\documentclass[pra,twocolumn,showpacs]{revtex4}
\usepackage{amssymb}
\usepackage{graphicx}% Include figure files
\usepackage{dcolumn}% Align table columns on decimal point
\usepackage{bm}% bold math
\usepackage{epsfig}
\newcommand{\beq}{\begin{equation}}
\newcommand{\eeq}{\end{equation}}
\newcommand{\beqa}{\begin{eqnarray}}
\newcommand{\eeqa}{\end{eqnarray}}
\newcommand{\ba}{\begin{array}}
\newcommand{\ea}{\end{array}}
\usepackage{color}

\begin{document}
\title{
% PRA 
% Interaction-induced demixing and localization in a two-species quantum dimer model
% JPB 
% Delocalization effects, entanglement entropy 
% and spectral collapse of boson mixtures in a double well
%
% Localization effects of two bosonic species trapped in a double well
A continuous-variable approach to the spectral properties and quantum states
\\
of the two-component Bose-Hubbard dimer}

\author{
% G. Mazzarella$^1$, 
F. Lingua$^1$, and V. Penna$^1$}

\affiliation{
% $^1$ Dipartimento di Fisica e Astronomia Galileo Galilei and
% CNISM, Universit$\grave{a}$ di Padova, Via Marzolo 8, 35131 Padova, Italy\\
$^1$
Dipartimento di Scienza Applicata e Tecnologia and u.d.r. CNISM, Politecnico di Torino,
Corso Duca degli Abruzzi 24, I-10129 Torino, Italy}

\date{\today}
%
%\date{\today}

\begin{abstract}
A bosonic gas formed by two interacting species
trapped in a double-well potential features macroscopic localization effects when 
the interspecies interaction becomes sufficiently strong. A repulsive interaction 
spatially separates the species into different wells while an attractive interaction confines both 
species in the same well. 
% We exploit this phenomen to  
We perform a fully-analytic study of the transitions from the 
weak- to the strong-interaction 
regime by exploiting the semiclassical method in which boson populations are represented in 
terms of continuous variables. We find an explict description of low-energy eigenstates and 
spectrum in terms of the model parameters which includes the neighborhood of 
the transition point.  
To test the effectiveness of the continuous-variable method we compare its predictions
with the exact results found numerically. Numerical calculations 
confirm the spectral collapse evidenced by this method when the space localization
takes place.
%
% We investigate the model in the semiquantum picture recently applied
% to various BH-like models which provides the equations describing the boson populations
% in terms of the model parameters. Such equations well reproduce for $|W|$ large enough
% the transition from the uniform ground state (delocalization regime with mixed species) to a
% localized ground state with separated ($W>0$) or mixed ($W<0$) species.
% These predictions are corroborated by our numerical results. Our approach well 
% describes the dramatic change of the model dynamical algebra and of the relevant 
% energy spectrum.
%

\end{abstract}
\maketitle

\section{Introduction}

Many-boson systems described in the Bose-Hubbard picture are characterized by density-density
interactions whose nonlinear character determines an extraordinarily rich scenario of dynamical
behaviors and properties.
In this framework and among many interesting aspects, a large attention has been focused on
small-size bosonic lattices since they provide a fertile ground to investigate the quantum-classical
correspondence and the role of nonlinear interactions  \cite{rev1}-\cite{rev12}. 
While the semiclassical approaches \cite{sc1}-\cite{sc3} to this class of systems are generally
not problematic, their study at the purely-quantum level remains a considerably hard task and the 
diagonalization of quantum Hamiltonians mainly relies on the use of numerical techniques.
   
An effective analytical method which has allowed, in many situations, to circumvent this difficulty
consists in reformulating the dynamics of low-energy bosonic states in terms of continuous
variables (CV) which represent the quantum numbers of boson populations. Fock states are thus
transformed in wave functions depending on the CV while the energy-eigenvalue equation can be
reduced, in the low-energy regime, to the problem of a multidimensional harmonic oscillator. 

This scheme has found large application in the last two decades 
% The approximation in terms of CV has been directly effected on the energy-eigenvalue problem in 
for studying the spatial fragmentation \cite{cvp1} and the spectral properties \cite{cvp3}
of condensates trapped in a double-well potential, the critical behavior \cite{cvp4} 
and the dynamical phase transition \cite{cvp5}, \cite{cvp6} leading to the emergence of
localized ground states in attractive condensates, and the collapse and revival \cite{cvp7} 
of nonlinear tunneling in Bose-Hubbard (BH) chains. 

While the CV approximation can be directly carried out
on the energy-state eigenvalue problem to reduce it to a solvable differential equation
as in papers \cite{cvp1}-\cite{cvp7}, 
a simple but useful generalization of this method consists in the derivation
of an effective Hamiltonian associated with the original model. This has been used
to reduce the BH chain to a solvable phonon-like quadratic Hamiltonian \cite{cvp8}, 
and to show how the potential provided by the effective Hamiltonian  completely
determines the ground-state properties of the attractive BH trimer \cite{cvp9}
and of a gas of dipolar bosons in a four-well ring \cite{cvp10}.

In this paper we apply the CV method to reproduce the mechanism governing the 
spectral collapse of energy levels, a phenomenon which often marks critical phenomena
involving the transition to new dynamical regimes. 
This is the case for nonlinear BH-like
models but also for models describing matter-photon interactions
whose nonlinearity is inherent in the spinor form of their Schr\"odinger problem.
Several examples are known such as 
the transition to the super-radiant phase in the Dicke
model, exhibiting the emergence of a quasicontinuous spectrum \cite{Em},
and the interaction-induced spectral collapse characterizing the
two-photon quantum Rabi model \cite{Fe} in which the Hamiltonian becomes
unitarily equivalent to a noncompact generator of su(1,1) \cite{PeRa}. 

The same effect distinguishes as well the transition of single-depleted-well
states from stable to unstable regimes in the BH trimer \cite{rev10}, \cite{VP}, and the emergence from
the delocalization regime of a fully-localized ground state in a double-well
system (dimer) with two bosonic components \cite{LMP}.  
The dimer system involving binary mixtures has recently raised a considerable interest, and 
its dynamical stability \cite{xu}, different types of self-trapping solutions \cite{satjia},
the Rabi-Josephson dynamics \cite{mazz2011}, the low-energy quantum states \cite{citro},
and the interspecies entanglement properties \cite{mujal} have been investigated.
A more extensive discussion on the nonlinear dynamics of multicomponent systems described in terms of 
discrete nonlinear Schr\"odinger equations and of their modulational instability can be found in 
\cite{baizakov}-\cite{kevre}. 

In reference \cite{LMP}, the two-component BH dimer has been investigated and its exact spectrum
has been compared with the spectrum derived through a Bogoliubov-like scheme. The derivation of
the latter, however, revealed how the implementation of this semiclassical approximation strongly 
depends on the dynamical regime in which is performed. More specifically, different dynamical regimes
involve totally different sets of microscopic bosonic modes enabling the diagonalization process.
In addition, the complex structure of the energy eigenstates resulting from this process is such
that extracting the significant physical information often is a non 
trivial task.
 
In this paper, the CV method is shown to offer a unified effective scheme able to determine
the spectrum for any choice of the model parameters, and to supply a complete description of the
spectral collapse emerging in the transition from the weak to strong-interaction regime. 
The study of a model including the occurrence of a known 
critical phenomenon allows us to better test the effectiveness of this method,
a central aspect of this work. 

After deriving the effective Hamiltonian for the two-component dimer in terms of continuous 
variables and the relevant minimum-energy configurations, we apply the CV method to reconstruct 
the energy levels of the systems and the explicit expression of the corresponding eigenstates.
We demonstrate as well how this methodology effectively describes, in a fully analytic way, 
the mechanism of the transition (heralded by the spectral collapse)
from a ground state with delocalized boson populations to a ground state where boson populations 
become strongly localized. 

In Section II we review the CV method and derive the model Hamiltonian
for the two-component dimer Hamiltonian within this scheme. Section III is devoted to
solve the boson-population equations incorporating the information about the 
minimum-energy configurations. In Section IV, we reconstruct the spectrum and the 
eigenstates. Finally, Section V is devoted
to compare exact results, found numerically, with the spectrum and the eigenstates
derived through the CV method.

%%%%%%%%%%%%%%%%%%%%%%%%%%%%%%%%%%%%%%%%%%%%%%%%%%%%%%%%%%%%%%%%%%%%%%%%%%%%%%%%
%%%%%%%%%%%%%%%%%%%%%%%%%%%%%%%%%%%%%%%%%%%%%%%%%%%%%%%%%%%%%%%%%%%%%%%%%%%%%%%%
%%%%%%%%%%%%%%%%%%%%%%%%%%%%%%%%%%%%%%%%%%%%%%%%%%%%%%%%%%%%%%%%%%%%%%%%%%%%%%%%
%
\subsection{The 2-component dimer model}

Ultracold bosons trapped in two potential wells are well described  by
the two-mode BH Hamiltonian  
$$
H_a =
\frac{U_a}{2}\Bigl [ a_{L}^{\dagger} a_{L}^{\dagger} a_{L} a_{L}
+
a_{R}^+ a_{R}^+ a_{R} a_{R} \Bigr ]
-J_a \big( a_{L}^+ a_{R}+ a_{R}^+ a_{L} \big ),
% \qquad c= a, b\; ,
$$
where $L$ ($R$) refers to the left (right) well, and the boson operators
$a_L$, $a_L^+$, $a_R$, $a_R^+$ 
satisfy the standard commutator $[a_\sigma, a_\sigma^+]$ $=1$ with $\sigma= L,R$. 
Parameters $U_a$ and $J_a$ are the boson-boson interaction
and the hopping amplitude, respectively. 
% often called dimer model
%
In the presence of two interacting atomic species, the spatial modes become four, $a_L$, $a_R$,
and $b_L$, $b_R$, for the components $A$ and $B$, respectively. The microscopic dynamics
of the system is described by the two-species dimer Hamiltonian (TDH) defined on
a two-site lattice
\begin{equation}
\label{bhtwo}
\hat{H} ={H}_{a}+ {H}_{b} + W \big (  a^+_{L} {a}_{L} b^+_{L} b_{L}
+
a^+_{R} {a}_{R} b^+_{R} {b}_{R} \big)
%{H}_{ab}
%%%\nonumber
\end{equation}
where ${H}_{a}$ and ${H}_{b}$ are the single-species Hamiltonians and
%
% $$ {H}_{ab}= W \big (  a^+_{L} {a}_{L} b^+_{L} b_{L}
% + a^+_{R} {a}_{R} b^+_{R} {b}_{R} \big) \;  $$
%
the interspecies interaction $W$ describes the coupling of the two components.
The further hopping parameter $J_b$ and intraspecies interaction $U_b$ occur
in $H_b$ describing the second component.
% and the two hopping amplitude $J_a$ and $J_b$. 
Since the total boson numbers
$$
N_a = N_{aL} + N_{aR}, \quad
N_b = N_{bL} + N_{bR} ,
$$
($N_{ar} = a^+_{r} {a}_{r}$, $N_{br} = b^+_{r} {b}_{r}$, $r=L, R$) 
of each bosonic component are conserved quantities being $[H, N_a]$ $=[H, N_b]$ $=0$, the eigenvalues of 
$N_a$ and $N_b$ represent two further significant parameters. We shall denote the
boson numbers of the two species with the same symbols $N_a$ and $N_b$ of their
number operators.

%%%%%%%%%%%%%%%%%%%%%%%%%%%%%%%%%%%%%%%%%%%%%%%%%%%%%%%%%%%%%%%%%%%%%%
% zzzz
\section{The continuous-variable method}
\label{cvp}

A useful description of the low-energy scenario of multimode bosonic models 
can be obtained by observing that physical quantities depending on the local 
populations $n_i$
(the eigenvalues of number operators $\hat {n}_i= \hat{c}^{\dagger}_i \hat{c}_i$)
can be reformulated in terms of continuous variables $x_i = n_i /N$
representing local densities \cite{cvp1}-\cite{cvp4}.
For boson number $N = \sum_i n_i$ large enough, Fock states
$|{\vec n } \rangle = |n_1, n_2, ..., n_L \rangle \equiv |x_1, x_2, ..., x_L \rangle $,
can be interpreted as functions of variables $x_i$ and creation/destruction
processes $n_i \to n_i \pm 1$ correspond to small
variations $|x_1, ..., x_i \pm \epsilon ,... ,x_L \rangle$ of state
$|x_1, ..., x_i ,... ,x_L \rangle$, where $\epsilon = 1/N \ll 1$.
Such an approach, in addition to simplify the energy-eigenvalue problem
associated to a multimode Hamiltonian ${H}$, also leads to a new effective Hamiltonian
written in terms of coordinates $x_i$ and of the corresponding generalized momenta
\cite{cvp9}. A well-known example \cite{cvp10} is provided by the BH Hamiltonian 
defined on a one-dimensional lattice
$$
\hat{H} = \frac{U}{2} {\sum}^M_{i=1}  \hat{n}_i (\hat{n}_i-1)
-J{\sum}_{rs} A_{rs} \hat{c}^{\dagger}_r \hat{c}_{s}\, ,
%\label{bhh}
$$
where $M$ is the lattice-site number, $r, s \in [1,M]$
and the adjacency matrix $A_{r s}$
is equal to $1$ for $s= r\pm 1$ and zero in the other cases.
By expanding up to the second order the quantity $\hat{H} |E \rangle$ 
in the corresponding eigenvalue problem $\hat{H} |E \rangle = E |E \rangle $, 
the latter takes the CVP form
\beq
( -D + V \, )\, \psi_E ({\vec x})
 = \, {E}  \, \psi_E  ({\vec x}) \, , 
% \qquad {\bar E}= {E}/({N^2 U})\, ,
\label{cvp1}
\eeq
including the generalized Laplacian
$$
D = \, N^2 U \tau  \sum_{rs}
\frac{\epsilon^2 }{2} A_{r s} \, \Bigl ( \partial_r -\partial_s \Bigr )\, {\sqrt { x_r \, x_s }}
\Bigl ( \partial_r -\partial_s \Bigr )\, ,
$$
with $\tau = J/(N U)$, and the effective potential
$$
V=N^2 U 
\sum^M_{r=1}  \left ( \frac{1}{2} x_r (x_r  - \epsilon)
- 2\tau  {\sqrt { x_r \, x_{r+1} }} \right )\, .
$$
The solutions $\psi_E ({\vec x})$ to problem (\ref{cvp1}) is easily found
by considering
the eigenvalues $E$ close to the extremal points (minima and maxima) of $V$
where the latter can be reduced to a quadratic form, namely,
to a multidimensional harmonic oscillator. Once $\psi_E ({\vec x})$ has been determined,
the eigenstates of the original eigenvalue problem for $\hat{H}$
are found to be $| E \rangle = \sum_{\vec x} \psi_E ({\vec x}) |{\vec x}\rangle $.
At the operational level, in addition to obtain an approximation
of the energy spectrum which seems to be effective (this aspect has been explored 
in Ref. \cite{cvp9} for the attractive BH model), one can exploit potential $V$ to 
obtain significant information about the ground-state configurations and its characteristic
regimes when the model parameters are varied. In the sequel, we focus our attention on $V$
and on the relevant extremal-point equations $\partial V/\partial {x_i} =0$ which allow to
determine at each lattice site the boson populations characterizing the ground 
state.

%%%%%%%%%%%%%%%%%%%%%%%%%%%%%%%%%%%%%%%%%%%%%%%%%%%%%%%%%%%%%%%%%%%%%%%%%%%%%%
\subsection{ The TDH in the continuous-variable picture }
\label{cvpa}

The application of the CV method to the TDH defined by (\ref{bhtwo})
yields the new eigenvalue equation
\beq
{\cal H}\, \psi_E ({\vec x}, {\vec y}) = \, {E}  \,\psi_E ({\vec x}, {\vec y})
\; ,
\label{cvp2}
\eeq
where ${\vec x}= (x_R, x_L)$ and $ {\vec y}= (y_R, y_L)$ and
$x_k = n_k/ N_a$ and $y_k = m_k/ N_b$ with $k = L,R$ describe
the populations of species $a$ and $b$, respectively, 
% and $U_0$ defined few rows below. 
Concerning $N_a$ and $N_b$
one should recall that the total boson number $N_a = n_L+n_R$ and $N_b = m_L+ m_R$ of
the two species are conserved quantities.
$\cal H$ contains the generalized Laplacian
$D= D_x + D_y$ in which, in addition to
$$
D_x = \; N_a J_a \epsilon_a^2 
\Bigl (
% \frac{\partial}{\partial x_{L}} - \frac{\partial}{\partial x_{R}} 
\partial_{x_{L}} - \partial_{x_{R}} \Bigr )\,
{\sqrt { x_L \, x_{R} }}
\Bigl ( 
% \frac{\partial}{\partial x_{L}} -\frac{\partial}{\partial x_{R}} 
\partial_{x_{L}} - \partial_{x_{R}} \Bigr )
\, ,
$$
one must include $D_y$, due to the second component. $D_y$ is found by replacing $N_a U_a \epsilon_a^2$ 
with $N_b J_b \epsilon_b^2$ and $x$ with $y$, where $\epsilon_r = 1/N_r$ and $r= a, b$. 
Then ${\cal H}$ becomes
$$
{\cal H} = \tau_a \epsilon_a^2 
\Bigl ( \partial_{x_{L}} - \partial_{x_{R}}
% \frac{\partial}{\partial x_{L}} - \frac{\partial}{\partial x_{R}} 
\Bigr )\,
{\sqrt { x_L \, x_{R} }}
\Bigl ( \partial_{x_{L}} - \partial_{x_{R}}
% \frac{\partial}{\partial x_{L}} -\frac{\partial}{\partial x_{R}} 
\Bigr ) 
$$
$$
+
\tau_b \epsilon_b^2 
\Bigl ( \partial_{y_{L}} - \partial_{y_{R}}
% \frac{\partial}{\partial y_{L}} - \frac{\partial}{\partial y_{R}}
\Bigr )\,
{\sqrt { y_L \, y_{R} }}
\Bigl ( 
\partial_{x_{L}} - \partial_{x_{R}}
%\frac{\partial}{\partial y_{L}} -\frac{\partial}{\partial y_{R}}
\Bigr ) + \; V
$$
whose potential $V$ has the form
$$
V =
-\gamma +
\frac{u_a}{2} \Bigl ( x^2_L + x^2_R \Bigr )
+\frac{u_b }{2} \Bigl ( y^2_L  + y^2_R  \Bigr ) 
$$
$$
+ w ( x_{L} y_{L} +x_{R} y_{R})
-2 \Bigl (\tau_a {\sqrt { x_R \, x_L }} + \tau_b {\sqrt { y_R \, y_L }} \Bigr )
\; .
% \label{pot}
% \end{equation}
$$
%%%%%%%%%%%%%%%%%%%%%%%%%%%%%%%%%%%%%%%%%%%%%%%%%%%%%%%%%%%
In $V$ the new parameters $\gamma = (U_a N_a+U_bN_b)/2$ and
\begin{equation}
w= W N_a N_b\; , \quad 
%
% U_a N_a^2 = U (U_a/U) f_a^2 N^2 = U N^2 u_a  --->  u_a = (U_a/U) f_a^2
%
\tau_k = J_k N_k \; , \quad  u_k= N^2_k U_k \; ,
\label{defin}
\end{equation}
%
% and because they simply shift $H$ with a constant term $u_a N_a+u_b N_b$.
%
with $k=a, b$, have been used. The conservation of boson populations $N_a$ and $N_b$, 
represented by equations $1 = x_R +x_L$, and $1 = y_R +y_L$ implies that two 
of the four coordinates $x_i$ and $y_j$  can be seen as dependent variables.
By introducing the new population-imbalance variables
$x= x_L-x_R$ and $y= y_L-y_R$ the bosonic populations
are thus described by
$x_L = (1+x)/2$, $x_R = (1-x)/2$, $y_L = (1+y)/2$, and $y_R = (1-y)/2$
while the effective Hamiltonian (EH) takes the form
% for example, $x_L = 1-x_R$ and $y_L = 1-y_R$, one has
% $$
% V= \frac{u_a }{2} \Bigl ( 1- 2x_R  + 2x^2_R   \Bigr )
% +\frac{u_b }{2} \Bigl ( 1- 2y_R  + 2y^2_R  \Bigr )
% + u \Bigl  ( 1-x_R - y_R + 2 x_{R} y_{R} \Bigr )
% - 2 \Bigl (\tau_a {\sqrt { x_R (1-x_R) }} + \tau_b {\sqrt { y_R (1-y_R) }} \Bigr )\, .
% $$ 
$$
{\cal H} = -D -\gamma + 
\frac{u_a }{4} \Bigl ( 1  + x^2 \Bigr )
+
\frac{u_b }{4} \Bigl ( 1  + y^2 \Bigr )
$$
\begin{equation}
+
\frac{w}{2} (1+ x y )
-
\Bigl (\tau_a {\sqrt { 1- x^2 }} + \tau_b {\sqrt { 1- y^2}} \Bigr ) \; .
\label{Hf}
\end{equation}
with the Laplacian
$$
D \simeq  2 \tau_a \epsilon_a^2 \sqrt{1- {\bar x}^2} \frac{\partial^2}{\partial x^2}
+
2 \tau_b \epsilon_b^2 \sqrt{1- {\bar y}^2} \frac{\partial^2}{\partial y^2} \; .
$$
The operator $D$ has been approximated by introducing
the quantities ${\bar x}$ and ${\bar y}$ representing the values of $x$ and $y$ 
for which $V$ reachs one of its extremal values and the EH essentially reduces to
a model of coupled harmonic oscillators.
%%%%%%%%%%%%%%%%%%%%%%%%%%%%%%%%%
%

\subsection{ The semiclassical  picture of TDH}
\label{cvpa}

It is interesting to highlight the link of TDH reduced to the form (\ref{Hf}) 
with the semiclassical version of TDH which exhibits, as the most part of multimode
boson models, a dynamics typically described  by discrete nonlinear Schr\"odinger equations 
\cite{kevre}. The semiclassical picture, in which boson operators $a_r$ and $b_r$ are 
replaced by local order parameters $ \alpha_r$, and $\beta_r$ ($r=$ $ L, R$), and
the semiclassical Hamiltonian $H_s$ associated to (\ref{bhtwo}) are discussed in 
Appendix \ref{semiTDH}. $H_s$ takes the form
(\ref{Hsm})
$$
{H}_s =
\frac{u_a }{4} \Bigl ( 1  + x^2 \Bigr )
+\frac{u_b }{4} \Bigl ( 1  + y^2 \Bigr )+ \frac{w}{2} (1+ x y )
$$
$$
- \Bigl (\tau_a {\sqrt { 1- x^2 }} \cos (2\theta_x ) + \tau_b {\sqrt { 1- y^2}}  \cos (2\theta_y ) \Bigr ) \; ,
%\label{Hsm}
$$
where  $x= (|\alpha_L|^2 -|\alpha_R|^2)/N_a$, $y = (| \beta_L|^2-|\beta_R|^2)/N_b$
are imbalance variables, and $\theta_x$,  $\theta_y$ the relevant canonically-conjugate angle variables
(see Appendix \ref{semiTDH}). 
The Hamilton equations are given by  ${\dot x} = \{x, H_s \}$, ${\dot x} = \{x, H_s \}$, and, in the specific case of 
$\theta_x$ and $\theta_y$, by the formulas
$$
\hbar N_a {\dot \theta}_x = \frac{wy }{2} +  \frac{u_a x }{2} + \frac{x \tau_a \cos(2\theta_x) }{ {\sqrt { 1- x^2 }}} \; , 
%\label{eqx}
$$
$$
\hbar N_b {\dot \theta}_y =  \frac{wx }{2} +  \frac{u_b y }{2} + \frac{y \tau_b \cos(2\theta_y)  }{ {\sqrt { 1- y^2 }}} \; .
%\label{eqy}
$$
The calculation of the minimum-energy states requiring that $\theta_x = \theta_y =0$ shows that such
equations exactly reproduces equations (\ref{geq}), discussed in the next Section, determining the 
extremal points of $V$. The search of the minimum-energy configurations thus appears to be closely 
related to imposing the stationarity condition for $V$, a key intermediate step in the CV method.

%%%%%%%%%%%%%%%%%%%%%%%%%%%%%%%%%%%%%%%%%%%%%%%%%%%%%%%%%%%%%%%%%%%%%
%%%%%%%%%%%%%%%%%%%%%%%%%%%%%%%%%%%%%%%%%%%%%%%%%%%%%%%%%%%%%%%%%%
\section{Boson-population equations and ground-state configurations}

The minimum-energy configurations 
are obtained by imposing the stationarity conditions for the potential $V$, expressed by
equations $\partial V /\partial x =0$ and $\partial V /\partial y =0$.
These give the boson-population equations
\begin{equation}
w\; y = -u_a x - \frac{2x\tau_a }{ {\sqrt { 1- x^2 }}}
\; , \,\,
w\; x  = -u_b y -\frac{2y\tau_b}{ {\sqrt { 1- y^2 }}}\; .
\label{geq}
\end{equation}
The latter allows one to identify the entire set of configurations 
$(x, y)$ corresponding to the extremal values of $V= V(x,y)$ and, in
particular, the one describing the ground state. Determining the expressions
of $x$ and $y$, written in terms of the model parameters, allows one to
derive the spectrum of the EH.

\subsection{Symmetric solutions with $w>0$}

The distinctive feature of this case is represented by the
assumptions $u_a =u_b \equiv u$ and $\tau_a =\tau_b \equiv \tau$ leading to the simplified
system
\beq
w y = -u x- \frac{2x\tau}{ {\sqrt { 1- x^2 }}}
\; , \,\,
w x = -u y- \frac{2y\tau}{ {\sqrt { 1- y^2 }}} \; .
\label{sreq}
\eeq
We assume as well that both the effective interactions $w$ (interspecies) 
and $u$ (intraspecies) are repulsive.
% Note that the assumption $u_a =u_b$ and $\tau_a =\tau_b$ (see definition (\ref{def1})) 
% entails that
%
The symmetric form of equations (\ref{sreq}) implies that any solution
necessarily satisfies the condition $y=-x$. This property
allows one to solve the previous equations analytically. By setting
$y=-x$ one finds
$$
w x = +u x + \tau \frac{2x}{ {\sqrt { 1- x^2 }}}\, ,
$$
giving the three solutions
\beq
x_0= 0\, , \quad x_1= \pm \sqrt{1- \frac{4\tau^2}{(w-u)^2} }\, .
\label{S1}
\eeq
% w+u_*  = - \tau \frac{2}{ {\sqrt { 1- x^2 }}}
To identify the regime in which $x_0= 0= y_0$ is the ground state we consider
the second-order expansion of $V$ around this point by means of the coordinate
representation $x= (q+p)/\sqrt 2$ and $y= (q-p)/\sqrt 2$ in terms
of the local variables $q$ and $p$ (some details about this calculation are given in  
appendix A). From
$$
V \simeq \frac{w+u}{2} -2\tau-\gamma + \frac{u +2\tau +w}{4} \; q^2 +
\frac{u +2\tau-w}{4}\; p^2
$$
one evinces that $y_0= x_0 =0$ is the ground state only if
$$
u +2\tau > w \; ,
$$
namely, if interspecies interactions are weak enough.

In the opposite case, $u +2\tau < w$, the point $x_0=y_0= 0$ 
becomes a saddle point separating two symmetric minima. The exploration
of the parameter space is then completed by determining the quadratic approximation
of $V$ close to the two separated minima.
The expansion of potential $V$ around $y_1=-x_1$, with  
$x_1$ given by (\ref{S1}), can be effected by using the
local parametrization $x = x_1 +q$ and $y= y_1 +p$. The potential takes the form
$$
V \simeq 
%  V(x_1, y_1)
u-\gamma -\frac{2\tau^2}{w-u} +
\frac{1}{4} \left ( u +w +\frac{|w-u|^3}{4\tau^2} \right ) q^2
$$
\beq
\label{coherence2}
+ \frac{1}{4} \left ( u -w +\frac{|w-u|^3}{4\tau^2} \right ) p^2
\;,
\eeq
showing how the solution relevant to $x_1$, $y_1$ is an energy minimum if
$u-w +{|w-u|^3}/{\tau^2} > 0$. The latter condition reduces to $w > u+2\tau$
making it evident that the solutions associated with $x_1$ indeed represent
(symmetric) energy minima.
The double-minimum configuration then appears when the (effective) interspecies
interaction $w$ becomes sufficiently strong. For $w-u \to 2\tau$ the macroscopic 
{\it coalescence effect} takes place in which the solution $x_1$ collapses into the 
origin $x_0 =0$.

Summarizing, the weak-interaction regime features the ground-state solution
$x=y=0$ with a uniform distribution $x_L = x_R = 1/2$, and $y_L = y_R = 1/2$:
the two components are equally distributed in the two wells and thus
totally delocalized.
In the strong-interaction regime one finds three solutions, but $x_0=0$ must be
excluded. For $x_1 > 0$ one has the ground-state configurations
$x_L = y_R< x_R = y_L$ while $x_L = y_R  > x_R = y_L$ is found when $x_1 < 0$.
These confirm the effect
of separation of the two components that, for $w$ large enough, tend to occupy different
wells thereby resulting strongly localized.

%%%%%%%%%%%%%%%%%%%%%%%%%%%%%%%%%%%%%%%%%%%%%%%%%%%%%%%%%%%%%%%%%
\subsection{Symmetric case with $w <0$}

With an attractive (effective) interaction $w <0$ equations (\ref{sreq}) become
$$
|w| y = u x+ \frac{2x\tau}{ {\sqrt { 1- x^2 }}}
\; , \,\,
|w| x = u y+ \frac{2y\tau}{ {\sqrt { 1- y^2 }}} \; ,
$$
which entail the simple, but substantial, change that solutions must satisfy
the identity $x=y$ instead of $y=-x$ (as the repulsive case).
Then, in addition to solution $x_0'=y_0'=0$, one discovers that
the two non uniform solutions are given by
$$
x_1'= \pm \sqrt{1- \frac{4\tau^2}{(w+u)^2} }\, .
$$
The derivation of the quadratic approximation of $V$ in the proximity of points relevant to
such solutions (see appendix (\ref{attQV})) shows that $x_0'=y_0'=0$ and $x_1'=y_1'$ 
describe the minimum energy in the regimes
$$
|w| <\; u + 2\tau  \; ,\quad  |w|  > \; u + 2\tau \; ,
$$
respectively. In particular, while solution $x_0'=y_0'=0$ again entails
uniformly distributed and delocalized components as in the repulsive case,
solutions $x_1'=y_1'$ are associated to the boson-population distributions
\beq
\label{gs2}
x_L = y_L\;  < \; x_R = y_R \, , \quad x_L = y_L \; >\; x_R = y_R\; ,
\eeq
showing how, for a sufficiently strong $|w|$, the two components with attractive interaction
tend to share the same well thus describing populations localized and mixed.

%%%%%%%%%%%%%%%%%%%%%%%%%%%%%%%%%%%%%%%%%%%%%%%%%%%%%%%%%%%%%%%%%%%%%%%%
\subsection{Some remarks}

The symmetric case includes the situation when the system is formed by
twin species.
In this special case the fact that $J_a=J_b$, $U_a=U_b$ and 
$N_a=N_b$ implicitly
entails that conditions $\tau_a=\tau_b$ and $u_a=u_b$ are satisfied. Remarkably,
if the twin-species assumption is relaxed, it is still possible to describe, within
the current symmetric-solution case, infinitely-many situations corresponding
to different choices of $N_k$, $W$, $U_k$ and $J_k$. To this end
it is sufficient to vary such parameters without violating
the constraints $w= W N_a N_b=  {\rm constant}$ and
\begin{equation}
% f^2_a \frac{U_a}{U} = f^2_b \frac{U_b}{U}\, ,\quad
% \frac{J_a f_a }{UN} = \frac{J_b f_b }{UN}\, ,
N^2_a U_a = N^2_b U_b\, ,\quad J_a N_a  = J_b N_b \; ,
\label{sym}
\end{equation}
entailing the two identities $u_a= u_b$ and $\tau_b = \tau_a$.
We conclude by noting how, in the case when $u_a \ne u_b$ and $\tau_a \ne \tau_b$,
no analytic approach is able to provide the explicit solutions of equations (\ref{geq}),
which must be found numerically. Simulations where slight deviations from the symmetric
case are assumed show that no substantial differences are found in the
minimum-energy scenario. 
With reference to the twin-species case mentioned above, in the following
we shall associate the case with {\it strong} and {\it weak} interactions 
to inequalities $w > u +2\tau$ and $w < u +2\tau$, respectively. Formula
$$
W= 4J/N + U,
$$
describes the critical condition $w = u +2\tau$ in term of 
$J_a=J_b \equiv J$, $U_a=U_b\equiv U$ and $N_a=N_b=N/2$. 

%%%%%%%%%%%%%%%%%%%%%%%%%%%%%%%%%%%%%%%%%%%%%%%%%%%%%%%%%%%%%%%%%%%%%%%%%%%%%%
%%%%%%%%%%%%%%%%%%%%%%%%%%%%%%%%%%%%%%%%%%%%%%%%%%%%%%%%%%%%%%%%%%%%%%%%%%%%%%
%%%%%%%%%%%%%%%%%%%%%%%%%%%%%%%%%%%%%%%%%%%%%%%%%%%%%%%%%%%%%%%%%%%%%%%%%%%%%%

\section{Spectrum and eigenstates}

{\it Weak repulsive interaction $W$}.
In this regime, characterized by $w < u +2\tau $, the minimum corresponds to $x_0= y_0 =0$ 
in the twin-species case. Then variables $x$ and $y$ of EF (\ref{Hf}) represent
the natural coordinates for obtaining its quadratic approximation close to the potential
minimum. By using the new variables $x= (q+p)/\sqrt 2$, $y= (q-p)/\sqrt 2$ in the
quadratic approximation of the EH, one finds
\beq
{\cal H} \simeq K
-2 \tau \epsilon^2 \Delta_{qp} + \frac{w+u +2\tau}{4}  q^2 +\frac{u -w +2\tau }{4} p^2 
\label{HQ1}
\eeq
with $\Delta_{qp} = {\partial_q^2} + {\partial_p^2}$, and $K= -\gamma-2\tau +{(w+u)}/{2}$.
%
%%%%%%%%%%%%%%%%%%%%%%%%%%%%%%%%%%%%%%%%%%%%%%%%%%%%%%%%%%%%
%%%%%%%%%%%%%%%%%%%%%%%%%%%%%%%%%%%%%%%%%%%%%%%%%%%%%%%%%%%%%%%%%%%%%%%%%%%%%%%

%%%%%%%%%%%%%%%%%%%%%%%%%%%%%%%%%%%%%%%%%%%%%%%%%%%%%%%%%%%%%%%%%%%%%%%%%%%%%%%
%%%%%%%%%%%%%%%%%%%%%%%%%%%%%%%%%%%%%%%%%%%%%%%%%%%%%%%%%%%%%%%%%%%%%%%%%%%%%%%%%%%%%%%%%%%%%%%%%%
\begin{figure}[h]
\begin{center}
\includegraphics[clip,width=\columnwidth]{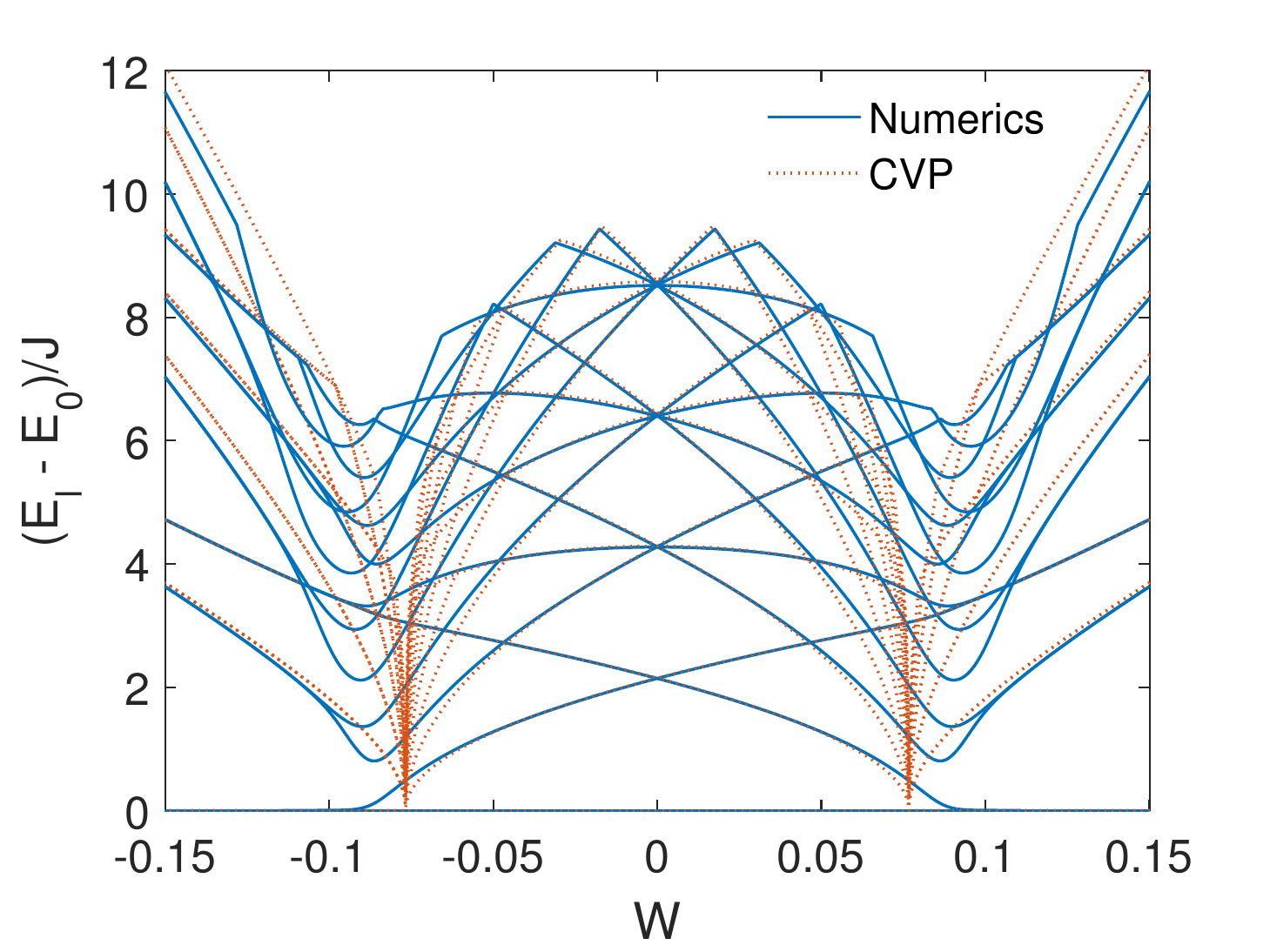}
\caption{(Color Online) First fifteen energy-levels as a function of interspecies interaction $W$
for intraspecies interaction $U =0.01$ (energy units in $J$) and total boson number $N=60$
with $N_a = N_b$. The plots compare numerical
results (continuous lines) with the analytical eigenvalues (dotted lines)
computed within the CV method.}
\label{figu1}
\end{center}
\end{figure}
%
%%%%%%%%%%%%%%%%%%%%%%%%%%%%%%%%%%%%%%%%%%%%%%%%%%%%%%%%%%%%\medskip \noindent
%%%%%%%%%%%%%%%%%%%%%%%%%%%%%%%%%%%%%%%%%%%%%%%%%%%
%
For twin boson populations $\epsilon_a =\epsilon_b$ so that $\epsilon = 2/N^2$.  
This harmonic-oscillator Hamiltonian
feature eigenvalues
$$
E_w (n ,m) = K
% \frac{w+u-2\tau}{2}
+\sqrt { 2\tau \epsilon^2 (u + 2\tau  + w ) } \Bigl ( n + {1}/{2} \Bigr )
$$
\beq
+  \sqrt { 2\tau \epsilon^2 (u + 2\tau - w ) } \; \Bigl ( m + 1/2 \Bigr )
\; ,
\label{EQ1}
\eeq
and the corresponding eigenstates are given by
\beq
\Psi_{n ,m}(q,p)
= \frac{
e^{- \frac{1}{2} ({q^2}/{\lambda^2} +{p^2}/{\nu^2}) } 
}{\sqrt{\pi \lambda \nu \; 2^{n +m} n! m! }} 
\; H_n \left (\frac{q}{\lambda} \right ) H_m \left (\frac{p}{\nu} \right ) 
\label{EF1}
\eeq
\medskip

\noindent
with $q= (x+y)/\sqrt 2$, $p= (x-y)/\sqrt 2$ and
$$
\lambda^2 = \sqrt{ \frac{8 \tau \epsilon^2}{w+u+2\tau} }\; ,\quad
\nu^2 = \sqrt{ \frac{8 \tau \epsilon^2}{u-w+2\tau} }\, .
$$
We note that the standard deviations $\lambda$ and $\nu$ controls
the extension of the gaussian factors in $\Psi_{n ,m}(q,p)$ and thus the
degree of localization of this state in the Fock space described (within
the CV method) by continuous variables $x$, $y$.
The amplitude of the quadratic approximation of $V$ contained in (\ref{HQ1})
essentially corresponds, at the minimum point, to
the gaussian curvature of $V$ which, in turn, is proportional to 
$1/(\nu \lambda)^4$.

The previous approximation is valid for weakly-excited states, namely, 
for energies relatively close to the ground-state energy. For the midspectrum states
the CV approach is no longer valid in that the assumption of continuity on which relies
may not hold \cite{cvp9}. 
%
%%%%%%%%%%%%%%%%%%%%%%%%%%%%%%%%%%%%%%%%%%%%%%%%%%%%%%%%%%%%%%%%
\begin{figure}[h]
\begin{center}
\includegraphics[width=0.7\columnwidth]{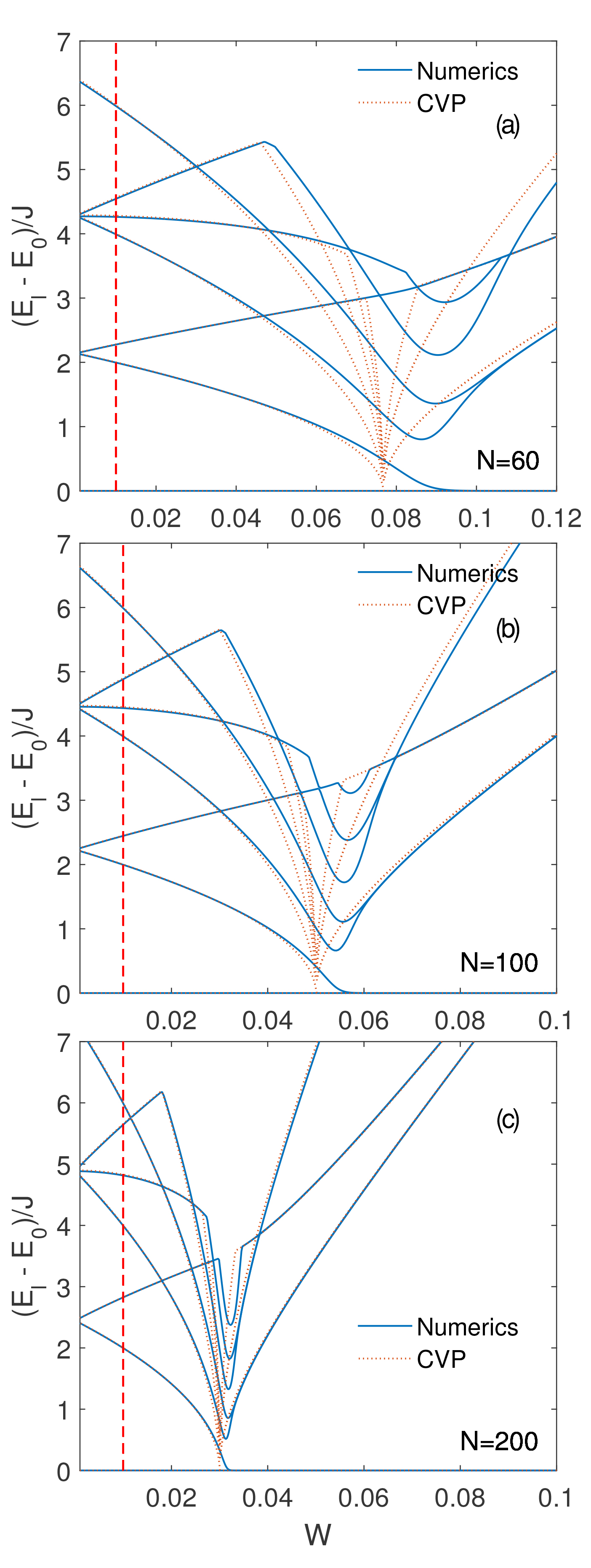}
\caption{(Color Online)
First seven energy-level as a function of interspecies interaction $W$ for $U =0.01$
(energy units in $J$) and boson number $N=60$ (panel (a)), $N=100$ (panel (b)) and
$N=200$ (panel (c)).
The plots compare numerical results (continuous lines) with the analytical eigenvalues
computed within the CV method (dotted lines). The vertical dashed line at $W=0.01$
shows the critical value ($\omega = u +2\tau$) where the transition takes
place in the thermodynamic limit.
}\label{figu2}
\end{center}
\end{figure}
%%%%%%%%%%%%%%%%%%%%%%%%%%%%%%%%%%%%%%%%%%%%%%%%%%%%%%%%%%%%%%%%%%%%

%%%%%%%%%%%%%%%%%%%%%%%%%%%%%%%%%%%%%%%%%%%%%%%%%%%%%%%%%%%%%%%%%%%%%%%%%%%%%%%%%%%%%%%%%
%%%%%%%%%%%%%%%%%%%%%%%%%%%%%%%%%%%%%%%%%%%%%%%%%%%%%%%%%%%%%%%%%%%%%%%%%%%%%%%%%%%%%%%%%
%%%%%%%%%%%%%%%%%%%%%%%%%%%%%%%%%%%%%%%%%%%%%%%%%%%%%%%%%%%%%%%%%%%%%%%%%%%%%%%%%%%%%%%%%

{\it Strong repulsive interaction $W$}. For $w > u +2\tau$, 
the single minimum of potential $V$ splits into two symmetric minima at 
$x= \pm |x_1|$ and $y= \mp |x_1|$. One easily calculates the quadratic approximation of 
EH (\ref{Hf}) in terms of the local-minima coordinates $\xi_x = x \pm |x_1|$ and $\xi_y = y\mp |x_1|$,
in which the double sign is referred to the two symmetric minima of $V$.
The further coordinate transformation $\xi_x,\xi_y \to q, p$ where
% $\xi_x= (q+p)/\sqrt 2$, $\xi_y= (q-p)/\sqrt 2$
$q = (\xi_x +\xi_y)/\sqrt 2$ and  $p = (\xi_x -\xi_y)/\sqrt 2$
leads to the diagonal, harmonic-oscillator form
$$
{\cal H}
\simeq  
- \frac{4 \tau^2 \epsilon^2}{|w-u|}\; \Delta_{qp}
+ 
\left ( \frac{u + w}{4} + \frac{(w-u)^3}{16 \tau^2} \right ) q^2 
%\Bigl ( \xi_x^2 +\xi_y^2 \Bigr )
%$$$$
$$
\beq
+\left ( \frac{u- w}{4} + \frac{(w-u)^3}{16\tau^2} \right ) p^2
+ u -\gamma -\frac{2\tau^2 }{w-u}\; ,
\label{HQ2}
\eeq
%
%%% $\sigma = u +\frac{|w-u|^3}{\tau^2} $
whose eigenvalues are given by
$$
E_s (n ,m) =
\epsilon (w-u) \sqrt {1 + \frac{4 \tau^2 (u + w)}{(w-u)^3} } 
\left ( n + \frac{1}{2} \right )
$$
\beq
+ \epsilon  \sqrt {(w-u)^2 - 4 \tau^2 }
\left (m+ \frac{1}{2} \right )+ u -\gamma -\frac{2\tau^2 }{w-u}
\; .
\label{EQ2}
\eeq
The corresponding eigenstates have the form
\beq
\Phi_{n ,m}(q,p)
= \frac{
e^{- \frac{1}{2} ({q^2}/{\lambda^2} +{p^2}/{\nu^2}) } 
}{\sqrt{\pi \lambda \nu \; 2^{n +m} n! m! }} 
\; H_n \left (\frac{q}{\lambda} \right ) H_m \left (\frac{p}{\nu} \right ) 
\label{EF2}
\eeq
with $q= (x+y)/\sqrt 2$, $p= (x-y \pm 2|x_1|)/\sqrt 2$, where the term $\pm 2|x_1|$ 
bears memory of the two symmetric minima of the current case, and
$$
\lambda^2 = \frac{8 \tau^2 \epsilon}{ \sqrt{(w-u) [4\tau^2 (w+u) +(w-u)^3]} }\; ,
$$
$$
\nu^2 = \frac{8 \tau^2 \epsilon}{ (w-u)\sqrt{ (w-u)^2 -4\tau^2} }\, .
$$
As in the weak-interaction case, such an approximation holds for weakly-excited states
and parameters $\nu$ and $\lambda$, related to the curvature of $V$, can be show
to control the localization character of these states in the Fock space. 
%
% namely, for energies relatively close to the ground-state energy. 
State (\ref{EF2}) actually corresponds to two independent states
associated to the same eigenvalue $E_s(n,m)$ which we denote with
\beq
\Phi^{\pm}_{n ,m}(x ,y) =\Phi_{n ,m} \left(  \frac{x+y}{\sqrt 2}, \frac{x-y \pm |x_1|}{\sqrt2} \right )  \; .
\eeq
The latter describe the low-energy eigenfunctions localized in the neighborhood of
the two minima of potential $V$. 
The degeneracy of the eigenvalues is a consequence of the partially-semiclassical character of 
the CV method. It can be removed by splitting each eigenvalue into a doublet 
$E^\pm_s(n,m) $ $=E_s(n,m) \pm \delta$,
where the splitting $\delta$ is obtained through the procedure described in \cite{landau} 
for the double-well potential. The simplest approximation of the eigenstates relevant to 
$E^r_s(n,m)$, $r = \pm$, is simply given by
\beq
\Psi^{r}_{n ,m}(x ,y) = %\frac{1}{\sqrt 2} 
\Bigl (\Phi^{+}_{n ,m} + r \Phi^{-}_{n ,m} \Bigr)/{\sqrt 2}
\; .
\label{psisymm}
\eeq 

%
%%%%%%%%%%%%%%%%%%%%%%%%%%%%%%%%%%%%%%%%%%%%%%%%%%%%%%%%%%%%%%%%%%%%%%%%%%%%%%%%%%%%%%%%%%%%%%%%%
%%%%%%%%%%%%%%%%%%%%%%%%%%%%%%%%%%%%%%%%%%%%%%%%%%%%%%%%%%%%%%%%%%%%%%%%%%%%%%%%%%%%%%%%%%%%%%%%%%
\begin{figure}[h]
\begin{center}
\includegraphics[clip,width=\columnwidth]{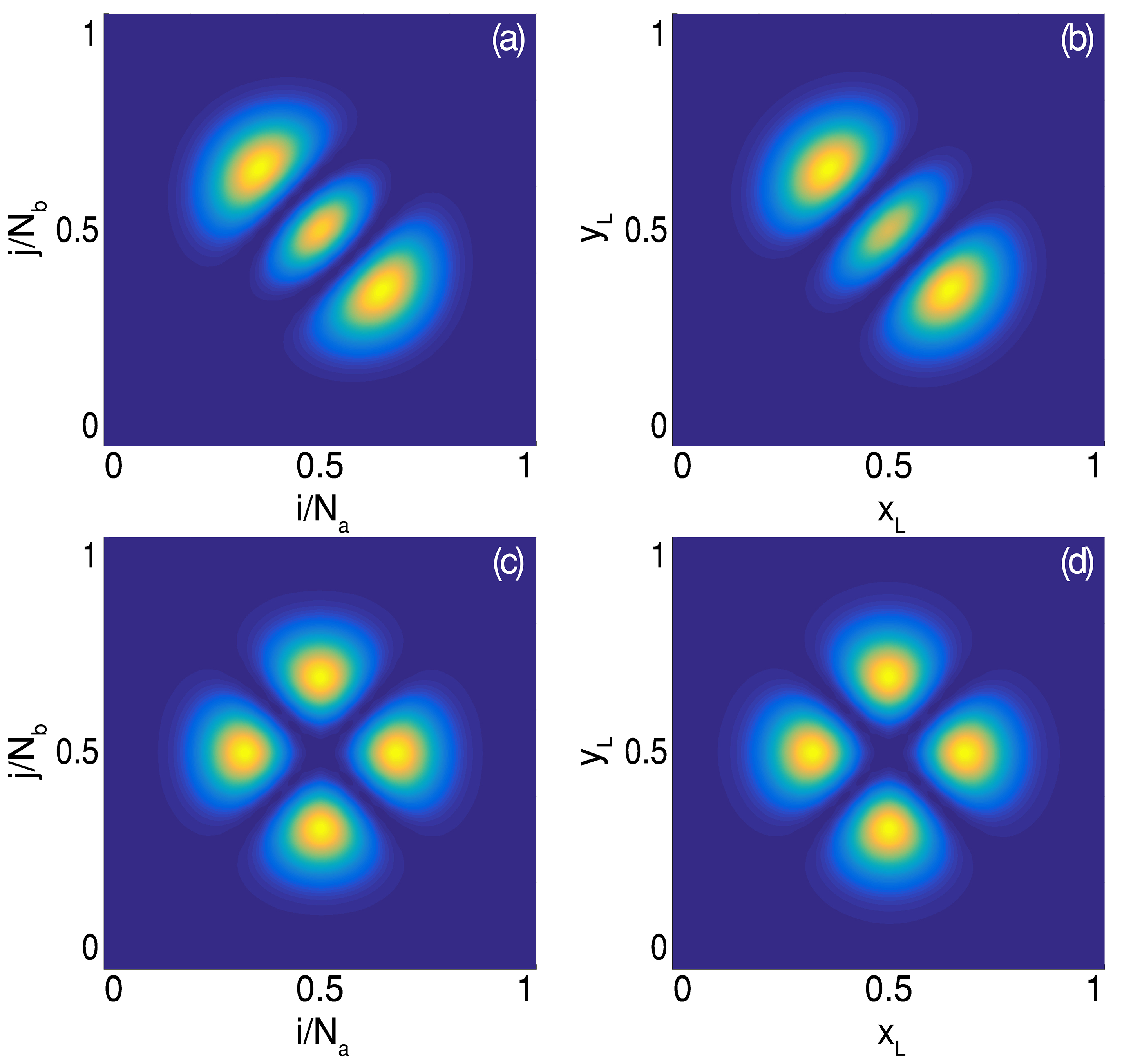}
\caption{
%(Color Online) Excited states $\Psi_{20}(q,p)$ (upper row, $l=3$) and
%$\Psi_{11}(q,p)$ (lower row, $l=4$) for $U/J= 0.01$, $W/J= 0.001$, $N_a=N_b =30$,
%calculated numerically (left panels) and within the CV method (right panels).
%These correspond to two of the three eigenvalues $E_l$, $l= 3, 4, 5$ forming
%the second plateau in Fig. \ref{figu4}.
(Color Online) Excited-state probability amplitudes $|c_{ij}^l|^2$ calculated numerically 
(panels (a),(c)), compared with the probability densities
$|\Psi_{n,m}|^2$ obtained by the CV method (panels (b),(d)).
The upper (lower) row concerns the excited states $\Psi_{2,0}(q,p)$ ($\Psi_{1,1}(q,p)$) 
for the energy level $l=3$ ($l=4$) for $U/J= 0.01$, $W/J= 0.001$, and $N_a=N_b =30$. These 
correspond to two of the three eigenvalues $E_l$, $l= 3, 4, 5$ forming
the second plateau in Fig. \ref{figu4}.
}
\label{figu3}
\end{center}
\end{figure}
%%%%%%%%%%%%%%%%%%%%%%%%%%%%%%%%%%%%%%%%%%%%%%%%%%%%%%%%%%%%%%%%%%%%%%%%%%%%%%%%%%

%%%%%%%%%%%%%%%%%%%%%%%%%%%%%%%%%%%%%%%%%%%%%%%%%%%%%%%%%%%%%%%%%%%%
\medskip

\noindent
{\it Attractive interspecies interaction}. In order to evidence the different
features characterizing the model with an attractive interaction we report
the eigenvalue spectra for $\omega <0 $.
These can be computed by following the same procedure of the repulsive case
$\omega >0$ (the corresponding Hamiltonians are shown in Appendix \ref{attQV}). 
For $|\omega|< u + 2\tau$ (weak interaction) one finds
$$
E_w' (n ,m)= K + \sqrt{2\tau\epsilon^2(u - |\omega| + 2\tau)}\Big(n + \frac{1}{2}\Big)
$$
\beq
+ \sqrt{2\tau\epsilon^2(u + |\omega| + 2\tau)}\Big(m + \frac{1}{2}\Big)
\label{EQ1p}
\eeq
where one should remember that $K ={(u +\omega )}/{2} -2\tau -\gamma$
and $\gamma = UN/2$ in the twin-component case.
For $|\omega|> u + 2\tau$ (strong interaction)
$$
E_s'(n, m)= 
\epsilon(|\omega| - u)\sqrt{1 + \frac{4\tau^2(|\omega| + u)}{(|\omega| - u)^3}}\Big(m + \frac{1}{2}\Big)
$$
\beq
+ \epsilon\sqrt{(|\omega| - u)^2 - 4\tau^2 } \Big( n + \frac{1}{2}\Big)
+u - |\omega| - \frac{2\tau^2}{|\omega| - u} - \gamma .
\label{EQ2p}
\eeq
\medskip

Figure \ref{figu1} well illustrates the perfect symmetry characterizing 
the energy spectrum when the interspecies interaction $w$ changes from positive (repulsive case) 
to negative (attractive case). 
This figure (and the subsequent ones) show the dependence of $(E_\ell -E_0)/J$ on $W/J$. 
Index $\ell$ in $E_\ell$ orders eigenvalues (\ref{EQ1}), (\ref{EQ2}), (\ref{EQ1p}), and (\ref{EQ2p}),  
according to their increasing values.
The reason for considering $E_\ell -E_0$ is that the eigenvalues $E_\ell$, $\ell \ge 0$ obtained with 
the CV method exhibit a finite shift with respect to the numerical eigenvalues. This deviation 
is a typical artifact of the quantization schemes including a semiclassical approximation \cite{gutz}. 
In the present case the deviations $E^{ap}_\ell - E^{ex}_\ell$ between approximate and exact 
eigenstates can be shown to be proportional to $1/N^2$ ($1/N$) in the weak (strong) interaction 
regime and thus to be negligible for $N$ large enough.

Figure \ref{figu1} compares the exact spectrum with the spectrum obtained through
the CV method for a total boson number $N =60$ and $N_a=N_b$. The critical points of the repulsive
and attractive cases are situated at $W/J \simeq + 0.076$ and $W/J \simeq - 0.076$, respectively.
At these values, both $E_w(n ,m)$, $E_s(n ,m)$ and $E_w'(n,m)$, $E_s'(n,m)$ tend to zero 
(see the dotted orange plots), while, in their proximity, the exact eigenvalues (blue continuous plots) 
exhibit a significant decrease culminating in a minumum.
Due to the relatively small value of $N$, the agreement between the exact an the approximate spectrum  
appears only at a sufficient distance from the critical points, but improves when $N$ is increased. 
This case is discussed in the next section where, owing to the spectrum symmetry, we focus on 
the case $W/J \ge 0$.

%%%%%%%%%%%%%%%%%%%%%%%%%%%%%%%%%%%%%%%%%%%%%%%%%%%%%%%%%%%%%%%%%%%%%%%%%%%%%%
%%%%%%%%%%%%%%%%%%%%%%%%%%%%%%%%%%%%%%%%%%%%%%%%%%%%%%%%%%%%%%%%%%%%%%%%%%%%%%
\section {Discussion}

We analyze the limit $w \to (u+2\tau)^{\pm}$. In this case, it is straightforward to check that 
the Hamiltonians (\ref{HQ1}) and (\ref{HQ2}) collapse into a unique one
$$
{\cal H}_S = {\cal H}_W
\simeq \, 
\left ( u -\tau \right ) - 2 \tau \epsilon^2
\left ( \frac{\partial^2}{\partial q^2} + \frac{\partial^2}{\partial p^2} \right )
+ \frac{u + 2\tau}{4} q^2 
\; ,
$$
in which the $p^2$-dependent terms go to zero due to the vanishing of
the frequencies $\sqrt{2\tau \epsilon^2 (u+ 2\tau- w)}$ in (\ref{HQ1}),
and $\sqrt{(u-w)^2 - 4\tau^2}$ in (\ref{HQ2}).
This effect causes in the eigenvalues (\ref{EQ1}) and (\ref{EQ2}) the 
spectral collapse, namely, the vanishing of the interlevel distance
relevant to the quantum number $m$ as shown by 

%
%%%%%%%%%%%%%%%%%%%%%%%%%%%%%%%%%%%%%%%%%%%%%%%%%%%%%%%%%%%%%%%%%%%%%%%%%%%%%%%%%%
\begin{figure}[h]
\begin{center}
\includegraphics[clip,width=\columnwidth]{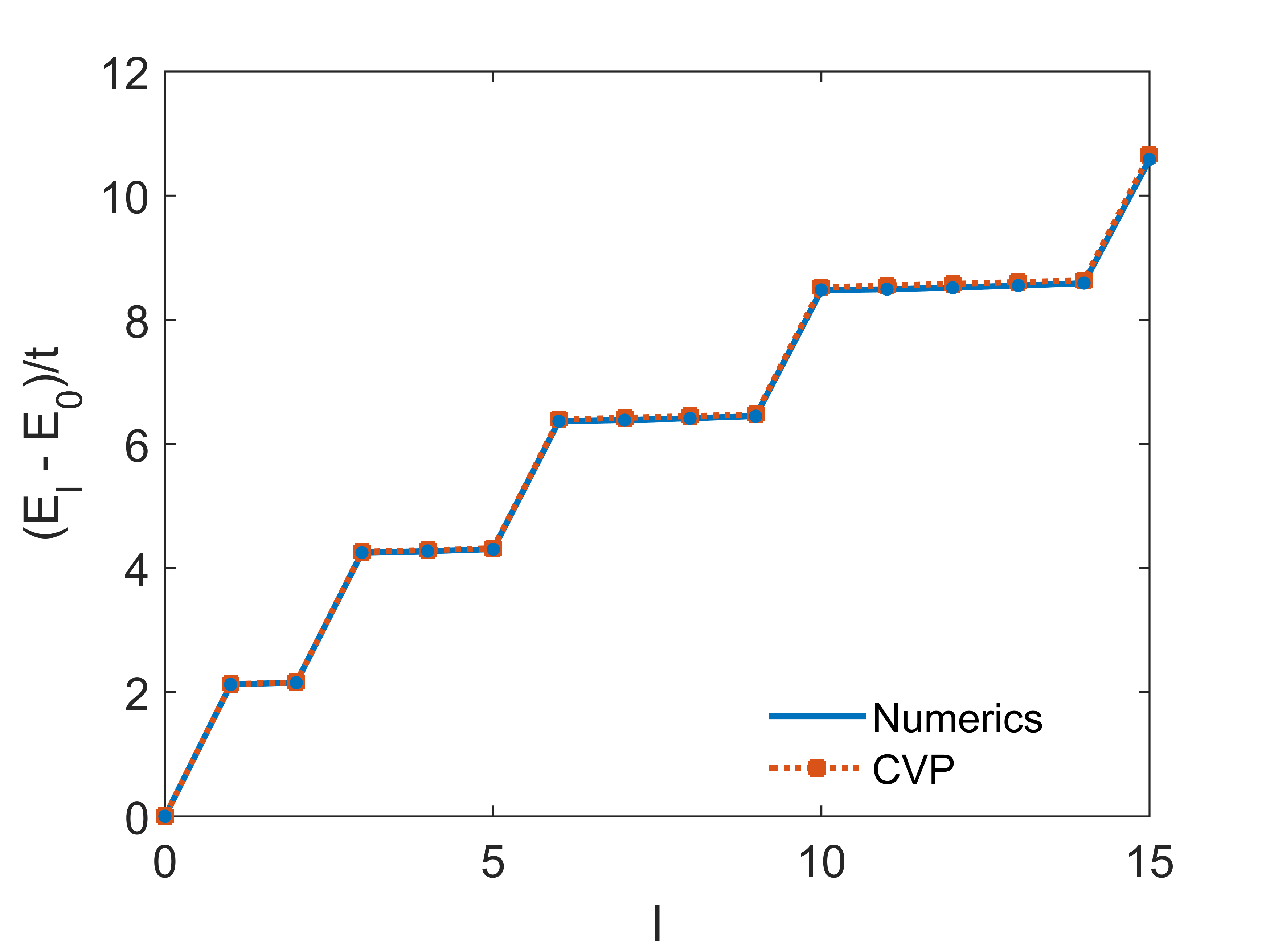}
\caption{(Color Online) Energy levels for $U/J= 0.01$, $W/J= 0.001$, $N_a=N_b =30$,
calculated numerically (continuous blue line) and within the CV picture (CVP).
The apparent formation of groups of degenerate eigenvalues (plateaux) is commented
in the text.}
\label{figu4}
\end{center}
\end{figure}
%%%%%%%%%%%%%%%%%%%%%%%%%%%\medskip \noindent

%
$$
E_w(n ,m) =  E_s( n ,m) \simeq  u-\gamma-\tau 
% \qquad \qquad
+ \epsilon \sqrt {  \tau (u+ 2\tau)} \left ( 2n+1 \right )
$$
\beq
%+ 2\epsilon \sqrt { \tau (u+ 2\tau)} \left ( n+ \frac{1}{2} \right )
+ \epsilon \sqrt {\tau |w- u- 2\tau| } \left (2m+1 \right )
\label{Elimit}
\eeq
for $w -u-2\tau \to 0$. When $w$ reachs the critical point $w \equiv u+2\tau$,
the free-particle term $- 2 \tau \epsilon^2 {\partial^2_p}$ in the Hamiltonian
entails the spectrum  
$$
E(n ,k) = u-\tau-\gamma
+ 2 \sqrt { \tau \epsilon^2 (u+ 2\tau)} \left ( n + \frac{1}{2} \right )
+ 2\tau \epsilon^2 k ,
$$
in which the contribution of quantum number $m$ is replaced by the $k$-dependent term,
while in
% $k$ being the eigenvalue of the plane-wave factor in the eigenstate
$$
\Phi_{n ,k}(q,p)
\propto
% \frac{1}{\sqrt{2 \pi \lambda \; 2^{n} n!}} 
e^{- {q^2}/{(2\lambda^2}) } H_n \left ({q}/{\lambda} \right ) e^{ik p}\; 
$$
the $p$-dependent gaussian becomes a plane wave.
% term whose standard deviation $\nu$ tends to infinity in both regimes. 
The progressive reduction of the interlevel distance 
(culminating, at the critical point, with the transition of the $m$-dependent energy band 
to a continuous energy distribution)
then represents the distinctive trait
marking the emergence of a ground state with a different 
structure. It is worth recalling that, this change consists in the transition from a 
ground state with two bosonic components totally mixed and delocalized ($w < u+2\tau$) to
a ground state whose components are completely localized ($w > u+2\tau$).
% 
%by construction, the assumption that boson populations are large enough.
The exact spectrum, determined by means of numerical simulations,
confirms the validity of the scenario emerging from the CV method
as soon as the boson numbers is sufficiently increased.

Figure \ref{figu2} describes the first seven energy levels as a function of interspecies 
interaction $W/J$ for total boson numbers $N=60, 100, 200$. The plots compare the eigenvalues
obtained numerically with the eigenvalues computed analytically by means of the CV method. 

At the critical point $W/J =$ $ U/J+ 4/N$ 
(derived from $w = u +2\tau$ thanks to the definitions (\ref{defin}) 
and populations $N_a=N_b =N/2$)
all the eigenvalues determined with the CV method continuously drop to zero. 
The vertical dashed line corresponds to the critical value of the interspecies interaction $W$
one finds in the thermodynamic limit $N \to \infty$ and with $U=0.01 J$ (energy units in $J$). 
In this limit one has $W=U$, reproducing the well-known critical value at which, for $W$ 
repulsive, the two components separate \cite{pra92}. 

Figure \ref{figu2} clearly shows how, by increasing $N$, the exact eigenvalues more and more 
tend to reproduce the critical behavior predicted by the CV method, while the critical value 
of $W/J$ approaches its limiting value 0.01.  We observe, however, that even for $N = 60$ the 
agreement between exact and CV-picture (CVP) spectrum becomes good right outside
the neighborhood of the critical point.

%%%%%%%%%%%%%%%%%%%%%%%%%%%%%%%%%%%%%%%%%%%%%%%%%%%%%%%%%%%%%%%%%%%%%%%%%%%%%%%%%%%%%%%%%%%%%%%%%
%%%%%%%%%%%%%%%%%%%%%%%%%%%%%%%%%%%%%%%%%%%%%%%%%%%%%%%%%%%%%%%%%%%%%%%%%%%%%%%%%%%%%%%%%%%%%%%%%%
\begin{figure}[h]
\begin{center}
\includegraphics[clip,width=\columnwidth]{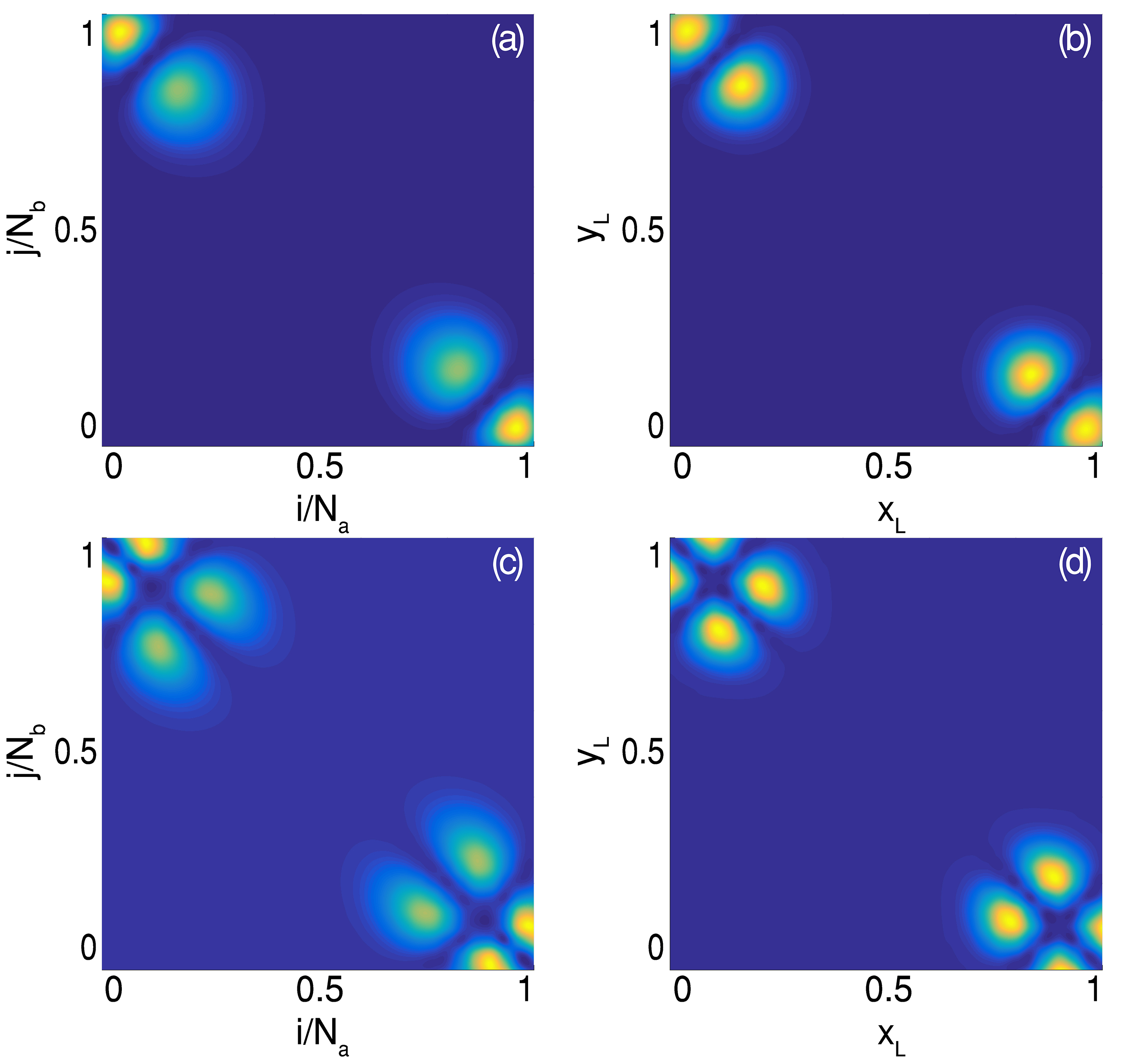}
\caption{
%(Color Online) Excited states $\Psi^+_{0,1}(x,y)$ (upper row, $l=2$) and
%$\Psi^+_{0,2}(x,y)$ (lower row, $l=8$) for $U/J= 0.01$, $W/J= 0.12$, $N_a=N_b =30$,
%calculated numerically (panels (a), (c)) and within the CV method (panels (b), (d)).
%These correspond to the eigenvalues $E_2= E_s(0,1)$ (second plateau),
%and $E_8= E_s(1,1)$ (fifth plateau) in Fig. \ref{figu6}.
(Color Online) Excited-state probability amplitudes $|c_{ij}^l|^2$ calculated numerically 
(panels (a),(c)), compared with the probability densities $|\Psi^+_{n,m}|^2$ obtained by 
the CV method (panels (b),(d)).
The upper (lower) row concerns the excited states $\Psi^+_{0,1}(x,y)$ ($\Psi^+_{0,2}(x,y)$) 
for the energy level $l=2$ ($l=8$) for $U/J= 0.01$, $W/J= 0.12$, and $N_a=N_b =30$. 
These correspond to the eigenvalues $E_2= E_s(0,1)$ (second plateau),
and $E_8= E_s(1,1)$ (fifth plateau) in Fig. \ref{figu6}.
}
\label{figu5}
\end{center}
\end{figure}
%
%%%%%%%%%%%%%%%%%%%%%%%%%%%%%%%%%%%%%%%%%%%%%%%%%%%%%%%%%%%%%%%%%%%%%\medskip \noindent
\subsection{Weakly-excited states}

We complete the comparison of the exact (numerical) scheme with the CV method
by considering the exact eigenstates and their CVP counterparts described by formulas 
(\ref{EF1}) and (\ref{psisymm}). 
The latter allow the reconstruction of the approximate eigenstates
$$
|\Psi_E \rangle = \sum_{{\vec x}, {\vec y}} \psi_E ({\vec x}, {\vec y}) |x_R, x_L, y_R, y_L \rangle
$$
according with formula (\ref{cvp2}), where the amplitude $\psi_E ({\vec x}, {\vec y})$
identifies with $\Psi_{nm} (q,p)$ or $\Psi^\pm_{nm} (x,y)$ 
(see formulas (\ref{EF1}) and (\ref{psisymm})) when $x_R$, $x_L$, $y_R$, $y_L$ are expressed  
in terms of variables $q$, $p$ (or $x$, $y$), and states $|x_R, x_L, y_R, y_L \rangle$ are the continuous 
form of Fock states $ |n_R, n_L, m_R, m_L \rangle$.

Figure \ref{figu3} illustrates the structure of some eigenstates in the weak-interaction 
regime $U/J= 0.01$, $W/J= 0.001$.
The probabilities $| c_{ij}|^2$ obtained from the exact eigenstates
$|E \rangle = \sum_i \sum_j c_{ij} (E) |N_a-i, i, N_b -j, j \rangle$ are compared with
their CVP counterparts $|\psi_{n, m}|^2$, where the amplitudes are 
$\psi_{n, m} (x_L, y_L) =$ $ \Psi_{n, m} (q, p)$
% In the weak-interaction regime one should remember that 
and $p =$ $ \sqrt 2 (x_L - y_L)$, $q = $ $\sqrt 2 (x_L + y_L -1)$.
One should remember that only two of the four coordinates $x_\alpha$, $y_\alpha$
are independent due to the constraints $x_R+ x_L= 1$, $y_R+ y_L=1$.

In Figure \ref{figu3}, the probability density of the eigenstates
associated to the three eigenvalues forming the second plateau of Fig. \ref{figu4}
are represented. In Figure \ref{figu3} and in the subsequent ones, dark blue stands for
a vanishing probability density while bright yellow denotes its relative maxima. 
Note that the presence of the energy plateaux shown in Fig. \ref{figu4} 
is only apparent: The groups of quasidegenerate eigenvalues with $E_l \simeq constant$ 
for $n+m = 0, 1, 2, ...$ are the consequence of the parameter choice $U/J= 0.01$$= 10\; W/J$ 
making the two harmonic-oscillator frequencies in (\ref{EQ1}) almost equal.

%
%%%%%%%%%%%%%%%%%%%%%%%%%%%%%%%%%%%%%%%%%%%%%%%%%%%%%%%%%%%%%%%%%%%%%%%%%%%%%%%%%%
\begin{figure}[h]
\begin{center}
\includegraphics[clip,width=\columnwidth]{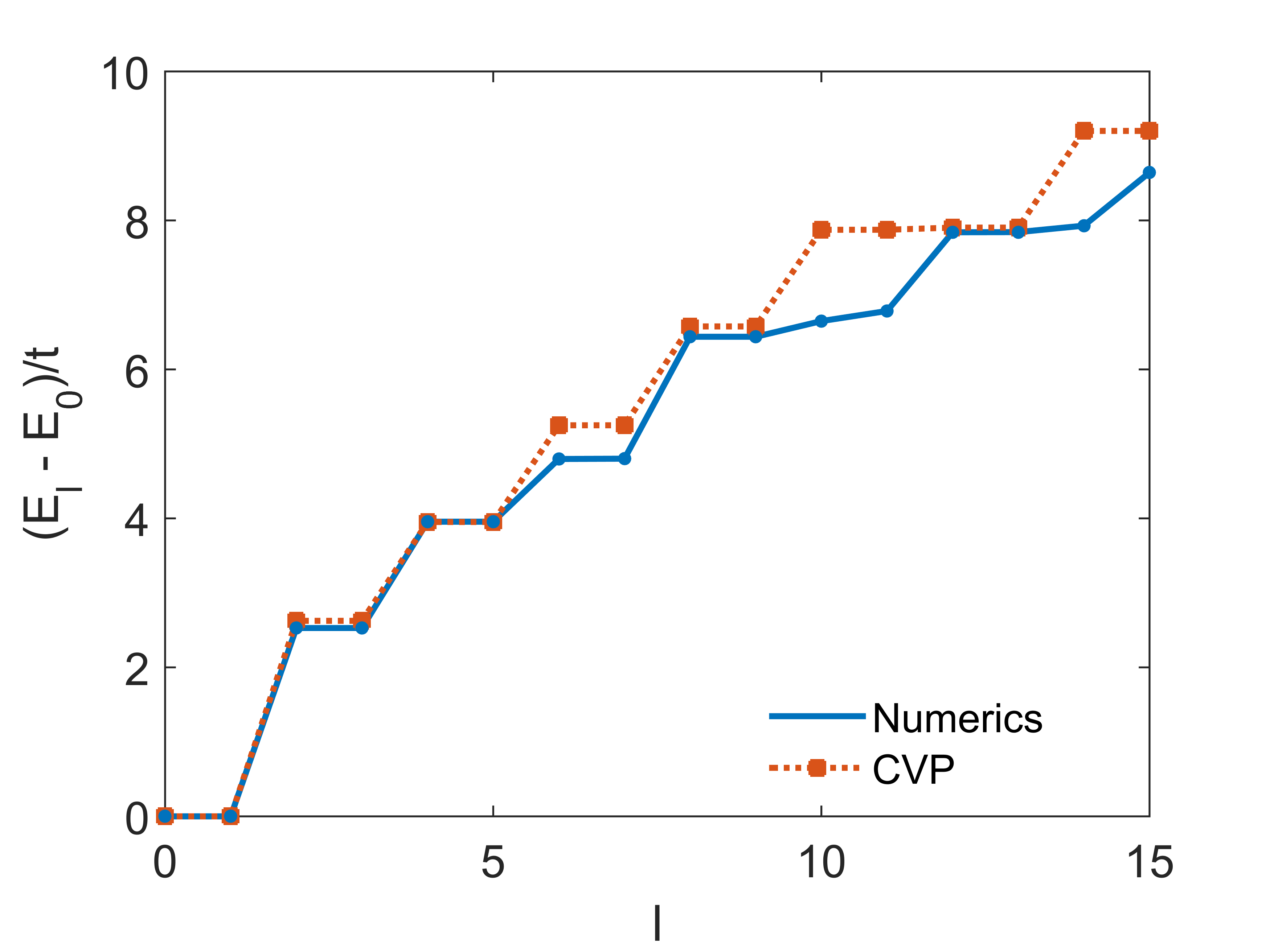}
\caption{(Color Online) Energy levels for $U/J= 0.01$, $W/J= 0.001$, $N_a=N_b =30$,
calculated numerically (continuous blue line) and within the CVP (orange dotted line).
The apparent formation of groups of degenerate exact eigenvalues (plateaux) is commented
in the text.}
\label{figu6}
\end{center}
\end{figure}
%%%%%%%%%%%%%%%%%%%%%%%%%%%\medskip \noindent
%%%%%%%%%%%%%%%%%%%%%%%%%%%%%%%%%%%%%%%%%%%%%%%%%%%%%%%%%%%%%%%%%%%%%%%%%%%%%%%%%%%%%%%%%%%%%%%%
%
%
%
%%%%%%%%%%%%%%%%%%%%%%%%%%%%%%%%%%%%%%%%%%%%%%%%%%%%%%%%%%%%%%%%%%%%%%%%%%%%%%%%%%%%%%%%%%%%%%%%%%
\begin{figure}[h]
\begin{center}
\includegraphics[clip,width=\columnwidth]{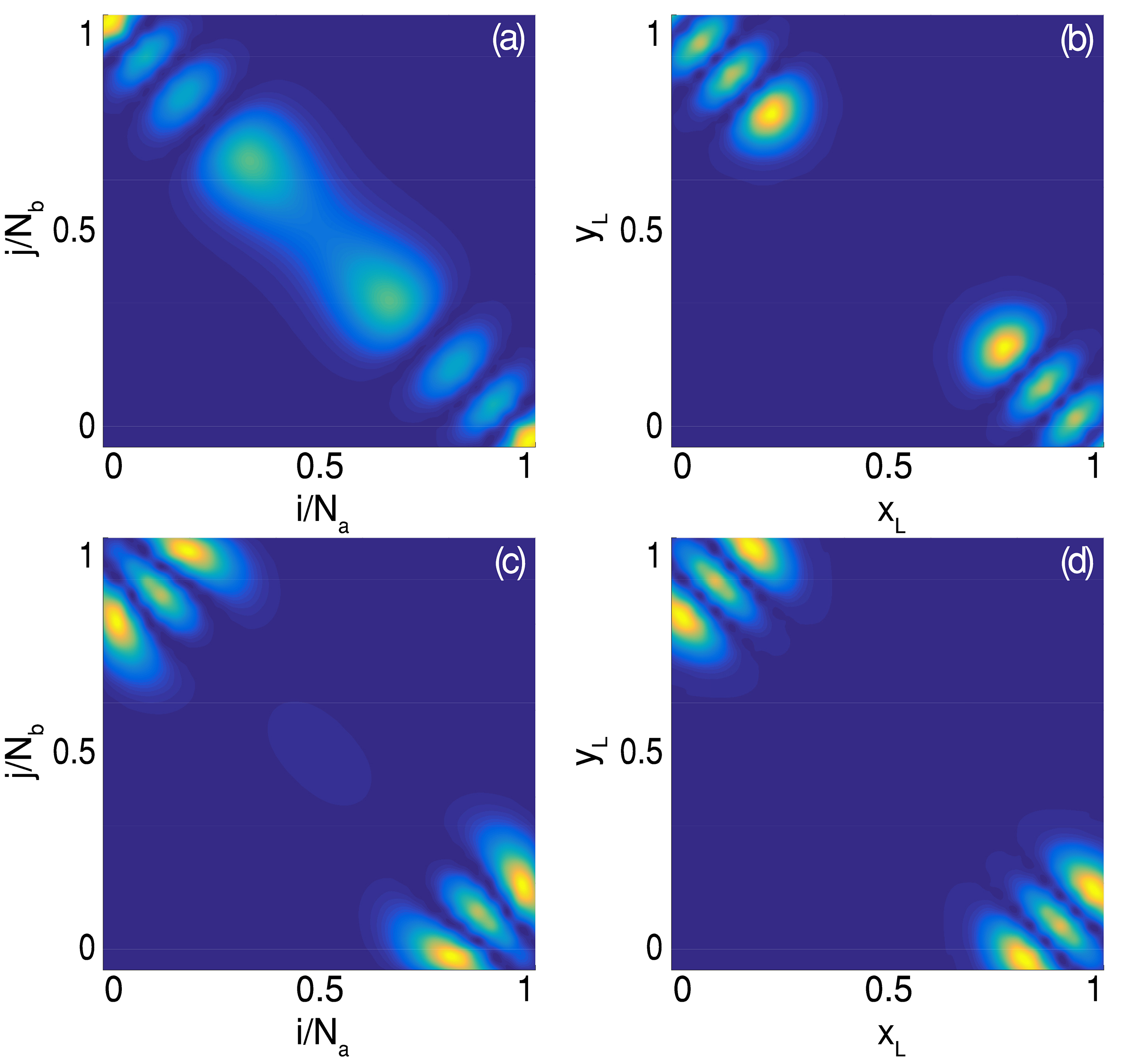}
\caption{
%(Color Online) Excited states $\Psi^+_{2,0}(x,y)$ (upper row, $l=10$) and
%$\Psi^+_{0,3}(x,y)$ (lower row, $l=12$) for $U/J= 0.01$, $W/J= 0.12$, $N_a=N_b =30$,
%calculated numerically (panels (a), (c)) and within the CVP (panels (b), (d)).
%These correspond the eigenvalues $E_{10}= E_s(2,0)$ and $E_{12}= E_s(0,3)$
%in Fig. \ref{figu6}.
(Color Online) Excited-state probability amplitudes $|c_{ij}^l|^2$ calculated numerically 
(panels (a),(c)), compared with the probability densities $|\Psi^+_{n,m}|^2$ obtained
by the CV method (panels (b),(d)).
The upper (lower) row concerns the excited states $\Psi^+_{2,0}(x,y)$ ($\Psi^+_{0,3}(x,y)$) 
for the energy level $l=10$ ($l=12$) for $U/J= 0.01$, $W/J= 0.12$, and $N_a=N_b =30$. 
These correspond to the eigenvalues $E_{10}= E_s(2,0)$ and $E_{12}= E_s(0,3)$
in Fig. \ref{figu6}.
}
\label{figu7}
\end{center}
\end{figure}
%
%%%%%%%%%%%%%%%%%%%%%%%%%%%\medskip \noindent

Figure \ref{figu3} displays the probability density of the eigenfunctions 
$\Psi_{2, 0}(q,p)$ and $\Psi_{1, 1}(q,p)$, which feature three and four peaks, respectively. 
The state $\Psi_{0, 2}(q,p)$ exhibits the same probability 
distribution (not shown) as $\Psi_{2, 0}(q,p)$ but the three peaks are placed along the 
second diagonal of the box.
This figure clearly shows how the exact and CVP probability densities are almost 
indistinguishable, a result further confirmed by other choices of $n$ and $m$.
Therefore the exact scheme and the CVP exhibit an excellent agreement. 
 
Figure \ref{figu5} displays the probability density of some eigenstates 
in the strong-interaction regime with $U/J= 0.01$, $W/J= 0.12$.
As in the weak-interaction case we compare
the $|c_{ij}|^2$ with the CVP probabilities $|\psi_{n, m}|^2$, but
in this regime $\psi_{n, m} (x_L, y_L) =$ $ \Psi^{\pm}_{n, m} (x, y)$,
the eigenfunctions (\ref{psisymm}) of energies $E_s(n,m)$.
Coordinates $x$ and $y$ are linear functions of $x_L$, $y_L$.
Figure \ref{figu5} compares the probability density for the excited states 
$\Psi^+_{01}(x,y)$ and $\Psi^+_{02}(x,y)$ (with energies $E_s(0,1)$ and $E_s(0,2)$, 
respectively) obtained in the CVP with those found in the exact scheme. These confirm 
the remarkable agreement of the CVP with numerical results.

The corresponding energy eigenvalues are illustrated in Figure \ref{figu6}.
The CVP eigenvalues are, by construction, degenerate and form the doublets 
$E_{2l} = E_{2l+1}$ (orange dots). The link with energies (\ref{EQ2}) 
is given by
$E_0 = E_s(0,0)$, $E_2 = E_s(0,1)$, $E_4 = E_s(1,0)$, $E_6 = E_s(0,2)$,
$E_8 = E_s(1,1)$ ... listed in increasing order.
It is worth remembering that this degeneracy is inherent in the CV method (see the 
discussion before eq. (\ref{psisymm})),
whereas the degeneracy of some exact eigenvalue is only apparent.

The important point concerning Fig. \ref{figu6} is that
at least ten CVP eigenvalues exhibit an excellent agreement with their numerical 
counterparts. Visible deviations appear in an intermittent way along the
eigenvalue sequence (see, for example, $E_6$, $E_7$, and $E_{10}$, $E_{11}$) but they
remain relatively small with respect to the trend of the the overall sequence.
The increase of boson number $N$ can be shown to reduces this effect.

Figure \ref{figu7} (upper panels) aims to illustrate the differences affecting the 
exact probability distribution and the CVP distribution for $\Psi^+_{20}(x,y)$, a state
whose eigenvalue $E_{10} = E_{s}(2,0)$ deviates from its numerical counterpart. 
Even if, in general, their overall structure is not too different, 
the upper left panel features two internal peaks exhibiting a weak separation, whereas, in the 
upper right panel, these peaks are completely separated.
Moreover, the left panel shows two major peaks (at the corners of the box) which
are almost negligible in the right panel. The two probability densities
again, almost perfectly, match to each other when considering the (non deviating) 
eigenvalue $E_{12} = E_{s}(0,3)$ relevant to the eigenstate $\Psi^+_{03}(x,y)$.

\begin{figure}[h]
\begin{center}
\includegraphics[clip,width=\columnwidth]{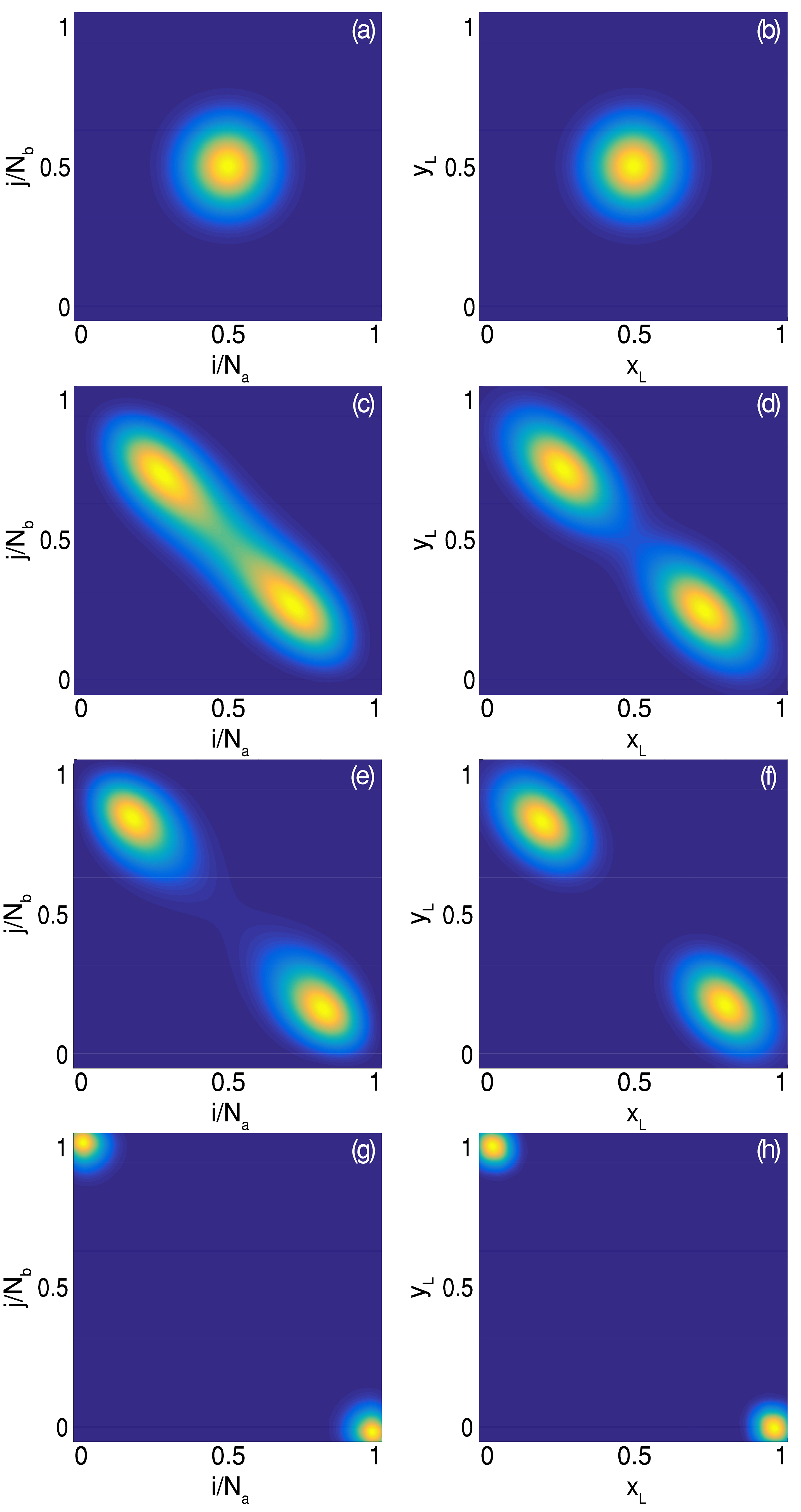}
\caption{
%(Color Online) Ground-state probability density for $U/J= 0.01$, $N_a= N_b =30$,
%and  (from top to bottom) $W/J= 0.001, 0.085, 0.093, 0.170$,
%calculated numerically (left column) and within the CV method (right column).
(Color Online) Probability densities of the ground state for $U/J= 0.01$, $N_a= N_b =30$,
and  (from top to bottom) $W/J= 0.001, 0.085, 0.093, 0.170$.
Right column: probability density obtained from states (\ref{EF1}) and (\ref{psisymm}) 
with $m=n=0$, within the CV method. Left column: probability amplitudes $|c_{ij}^0|^2$ for 
the exact ground state calculated numerically.
}
\label{figu8}
\end{center}
\end{figure}

We conclude by showing in Figure \ref{figu8} the sequence illustrating the 
probability densities of the ground state when $W/J$ ranges from the weak to the strong
interaction regime (up-to-bottom). For $W/J = 0.001$ a unique central peaks appears at $x_L = y_L = 0.5$
($\to x_R = y_R = 0.5$) meaning that the configuration with the maximum probability is that 
where the two components are equally distributed in the two wells. The boson populations 
are mixed and delocalized. For $W/J = 0.170$, the two peaks emerging from the transition
implies that $x_L \simeq 0$, $y_L \simeq 1$ and $x_R \simeq 1 0$, $y_R \simeq 0$, namely,
the two component are fully separated.
The agreement of numerical results and CVP predictions is quite satifactory.

%%%%%%%%%%%%%%%%%%%%%%%%%%%%%%%%%%%%%%%%%%%%%%%%%%%%%%%%%%%%%%%%%%%%%%%%%%%%%%%%%%%%%%%%%%%%%%%%%
%%%%%%%%%%%%%%%%%%%%%%%%%%%%%%%%%%%%%%%%%%%%%%%%%%%%%%%%%%%%%%%%%%%%%%%%%%%%%%%%%%%%%%%%%%%%%%%%%%
%
%%%%%%%%%%%%%%%%%%%%%%%%%%%%%%%%%%%%%%%%%%%%%%%%%%%%%%%%%%%%%%%%%%%%%
%
%

\section{Conclusions}

We have studied the effectiveness of the CV method by applying this scheme to the 
BH-like Hamiltonian describing a bosonic gas with two components trapped in a double-well 
potential. As it is well known, this system exhibits a macroscopic dynamical phase transition 
to states with localized populations when the effective interaction $W/U$ is large enough.
The presence of this transition plays an important role in our analysis since it makes 
the application of the CV method more demanding and thus more significant.  
We have analyzed the low energy spectrum and its eigenstates by considering both
the repulsive and the attractive regime of $W$. 

After reformulating, in Section II, the TDH within the 
continuous-variable picture, we have calculated the energy eigenvalues and the corresponding 
eigenstates in Section III. In this Section,
we have also showed that the reduction of the interlevel distance 
predicted by the CV method close to the transition point is confirmed 
by numerical simulations.
These also succeed in reproducing the spectral collapse for number of bosons sufficiently 
large, a condition which well fits the basic assumption of the CV method that the local 
population fractions $n_i /N$ are almost continuous.

To further check the effectiveness of the CV method we have compared both  
the weakly excited states and the corresponding energy levels
derived within the CV method with those determined numerically.
While in the weak interaction regime the agreement is excellent, in the strong 
interaction regime some eigenvalues exhibit visible but limited deviations
from their numerical counterparts. Such deviations appear in an intermittent way
in correspondence to sufficiently excited states 
and, rather reasonably, seem to be related to the intrinsic degeneracy of the CVP 
eigenvalues when the minimum of the potential splits into two separated minima. 

The agreement is again considerably good when comparing the probability density
of the exact and the CVP ground state both in the weak and in the strong interaction 
regime. In general, the CVP eigenstates closely mimic the exact eigenstates whenever
a numerical eigenvalue well matches the CVP eigenvalue.  

The previous analysis indeed suggests that the CV method provides an effective
approach for describing the energy spectrum and the eigenstates of multimode
bosonic systems. The discrepancies which partially affect the spectrum in certain 
regimes seem to have negliglible effects on the critical behavior leading to
the spectral collapse provided that a large number of bosons is involved. 
This is confirmed as well by the successful application of the CV method 
for detecting the critical properties of the self-trapping transition
in the attractive BH trimer \cite{cvp9}. 
The great feasibility of this method within multimode bosonic
systems promises a wide range of applications in the field of atomic currents 
\cite{atom1}-\cite{atom3} and more in general of atomtronics devices 
\cite{atom4}, \cite{atom5}.

Concluding, the effects discussed 
in this paper should be accessible to experimental observations  by confining mixtures in a double-well trap.
As is well known, the semiclassical dynamics of a single-component condensate has been successfully investigated
in a double-well device realized by \cite{albiez},  \cite{anker}, and has shown the nonlinear oscillations predicted
by the theory and the inherent self-trapping phenomenon.  
As in the single-component case, the double-well geometry should be realized by superposing the (sinusoidal) linear potential
confining mixtures \cite{lens}, \cite{gadway} in optical lattices  with a parabolic trap of controllabe amplitude. 
Further develpments in the dynamics of mixtures in multiwell systems 
are expected from the realization of the ring geometry designed in \cite{atom3}.
% Further develpments are expected concerning the realization of multiwell ring 
% systems according with the scheme designed in ...
%

%
%
% \acknowledgments
% GM and VP acknowledge Ministero dell'Istruzione, della Ricerca, e dell'Universit\`a (MIUR) 
% for the support provided by PRIN Grant No. 2010LLKJBX.
%%%%%%%%%%%%%%%%%%%%%%%%%%%%%%%%%%%%%%%%%%%%%%%%%%%%%%%%%%%%%%%%%%%%
%%%%%%%%%%%%%%%%%%%%%%%%%%%%%%%%%%%%%%%%%%%%%%%%%%%%%%%%%%%%%%%%%%%%

\begin{appendix}

%%%%%%%%%%%%%%%%%%%%%%%%%%%%%%%%%%%%%%%%%%%%%%%%%%%%%%%

\section{Semiclassical form of the TDH}
%%%%%%%%%%\label{attrapp}
\label{semiTDH}

The derivation of the semiclassical TDH can be performed by means of the coherent-state variational method 
where operators become classical variables within a sort of generalized 
Bogoliubov scheme \cite{sc1}.  The semiclassical Hamiltonian associated to (\ref{bhtwo}) is easily found to be
$$
H_s = H_a+ H_b + W (|\alpha_L|^2|\beta_L|^2+|\alpha_R|^2|\beta_R|^2),
$$
where $H_a = -J_a (\alpha^*_L \alpha_R +C. C.)+ U_a \sum_r |\alpha_r|^4/2 $ and $H_b$
has the same form with $\beta_r$ ($J_b$ and $U_b$) in place of $\alpha_r$  ($J_a$ and $U_a$). The classical quantities
$N_a = |\alpha_L|^2 +|\alpha_R|^2$ and $N_b =| \beta_L|^2+|\beta_R|^2$ can be shown to be conserved quantities as in the 
quantum picture. By using the classical version  $x= (|\alpha_L|^2 -|\alpha_R|^2)/N_a$ and $y = (| \beta_L|^2+|\beta_R|^2)/N_b$
of the operators leading to the EH (\ref{Hf}), one obtains, up to a constant term, 
$$
{H}_s =
\frac{u_a }{4} \Bigl ( 1  + x^2 \Bigr )
+\frac{u_b }{4} \Bigl ( 1  + y^2 \Bigr )+ \frac{w}{2} (1+ x y )
$$
\begin{equation}
-
\Bigl (\tau_a {\sqrt { 1- x^2 }} \cos (2\theta_x ) + \tau_b {\sqrt { 1- y^2}}  \cos (2\theta_y ) \Bigr ) \; ,
\label{Hsm}
\end{equation}
where $\theta_x = ( \phi_L - \phi_R) /2$, $\theta_y = ( \nu_L - \nu_R)/2$ are angle 
variables canonically conjugate with the action variables  $x$ and $y$ satisfying 
the Poisson Brackets $\{x, \theta_x \}= 1/(\hbar N_a)$, $\{y, \theta_y \}= 1/(\hbar N_b)$. 
Variables $\phi_r$ ($\nu_r$) are the phases of the local order parameters 
$\alpha_r = |\alpha_r| e^{i\phi_r}$
($\beta_r = |\beta_r| e^{i\nu_r}$). The Poisson brackets of $ |\alpha_r|^2$, $ |\beta_r|^2$ 
with $\phi_r$, $\nu_r$ can be easily evinced from the canonical ones 
$\{ \alpha_r, \alpha^*_r \}= 1/(i\hbar)$, $\{ \beta_r, \beta^*_r \}= 1/(i\hbar)$ 
supplied by the coherent-state variational method and reminescent of the boson mode commutators.
$[ a_r, a^+_r ]= 1$, $[ b_r, b^+_r ] = 1$. 

% Interestingly,  for $\theta_x = \theta _y =0$, Hamiltonian $H_s$ coincides with potential $V$ of the EH.

\section{Quadratic approximation of $V$ for $w<0$}
%%%%%%%%%%\label{attrapp}
\label{attQV}

We perform the quadratic approximation of $V$ in the proximity 
of its local minima, focusing on the attractive case (the same scheme holds in the
repulsive case). The minimum coordinates are given by
\begin{equation} 
x_0^\prime=y_0^\prime=0,\;\;\;\; x_1^\prime=\pm\sqrt{1 - {4\tau^2}/{(|\omega|-u)^2}} 
\; .
\end{equation}
By expanding potential $V$ in the proximity of its minima one finds 
that points $x_0^\prime=y_0^\prime=0$ and $x_1^\prime=y_1^\prime$ describe
the minimum-energy configuration 
in the regimes $|\omega|< u +2\tau$ and $|\omega|> u +2\tau$ respectively.

In the attractive case $\omega < 0$,
the EH is $\mathcal{H}= V-D$,
where $D$ has the same form as in (\ref{Hf}), 
%
% \begin{equation}
% D=2\tau\epsilon^2\sqrt{1 - \bar{x}^2}\frac{\partial^2}{\partial x^2} 
% + 2\tau\epsilon^2\sqrt{1 - \bar{y}^2}\frac{\partial^2}{\partial y^2}
% \end{equation}
%
and
$$
V=-\gamma + \frac{u}{4}\Big( 1 + x^2\Big) + \frac{u}{4}\Big( 1 + y^2\Big)
$$
\beq
-\frac{|\omega|}{2}\Big( 1 + xy\Big) - \tau\Big( \sqrt{1 + x^2} + \sqrt{1 + y^2}\Big)\; ,
\eeq
where $\gamma = UN/2$.
Potential $V$ is represented around the potential minima by means of its Taylor expansion.
%
% $$
% V= V(\bar{x},\bar{y}) + \frac{1}{2}\Big[\frac{\partial^2 V}{\partial x^2}\xi_x^2 
% + \frac{\partial^2 V}{\partial y^2}\xi_y^2 + 2\frac{\partial^2 V}{\partial x\partial y}\xi_x\xi_y\Big]
% $$
%
The resulting quadratic form is written in terms of local coordinates
$\xi_x=(x - \bar{x})$ and $\xi_y=(y - \bar{y})$ and $\bar{x}$, $\bar{y}$.
%
%%%%%%%%%%%%%%%%%%%%%%%%%%%%%%%%%%%%%%%%%%%%%%%%%

\emph{Weak interspecies interaction}. 
For $|\omega|< u + 2\tau$, the minimum coordinates
are $\bar{x}=x_0^\prime=0$ and $\bar{y}=y_0^\prime=0$. In this case
$$
\partial_x^2 V= \partial_y^2 V= \frac{u}{2} + \tau\; , \quad \partial_y \partial_x V=-\frac{|\omega|}{2}
\; .
$$
Setting $\xi_x=(q+p)/\sqrt{2}$ and $\xi_y=(q-p)/\sqrt{2}$ the EH
becomes ${\cal H} = - 2\tau \epsilon^2 (\partial^2_q + \partial^2_p) + V$
where the expanded potential reads
$$
V=K' + \frac{u - |\omega| + 2\tau}{4}q^2  + \frac{u + |\omega| + 2\tau}{4}p^2
$$
with 
$K' = ({u - |\omega| - 4\tau} - {UN})/{2}$.
Then the eigenvalues of the two independent harmonic oscillators occurring in $\cal H$  
can be easily computed. One finds
$$
E_W(n,m)=K' + \sqrt{2\tau\epsilon^2(u - |\omega| + 2\tau)}\Big(n + \frac{1}{2}\Big)
$$
\beq
+ \sqrt{2\tau\epsilon^2(u + |\omega| + 2\tau)}\Big(m + \frac{1}{2}\Big)
\eeq
\medskip

\noindent
\emph{Strong interspecies interaction}. 
For $|\omega|> u + 2\tau$, the coordinates of the potential minimum are $\bar{x}=x_1^\prime$ 
and $\bar{y}=y_1^\prime$. In this case
$$
\partial_x^2 V = \frac{u}{2} + \frac{\tau}{(1- x^2)^{\frac{3}{2}}}
\; ,\,\,\,
\partial_y^2 V = \frac{u}{2} + \frac{\tau}{(1- y^2)^{\frac{3}{2}}}
\; 
$$
and $\partial_x \partial_y V =-{|\omega|}/{2}$.
By setting $\xi_x=(q+p)/\sqrt{2}$ and $\xi_y=(q-p)/\sqrt{2}$, the expanded potential reduces to
\beq
V(q,p) = K'' + \frac{w_q^2}{2} \; q^2  + \frac{w_p^2}{2} \; p^2
\eeq
with
$$ 
w^2_q =  \frac{u - |\omega|}{2} + \frac{(|\omega| - u)^3}{8 \tau^2}
\; , \,
w^2_p = \frac{u + |\omega|}{2} + \frac{(|\omega| - u)^3}{8\tau^2},
$$
and
$K'' =  ({u - |\omega| - 4\tau} -{UN})/{2}$,
while the Hamiltonian of the system takes the form
\beq
\mathcal{H}= V_1^\prime - 
\frac{4\tau^2\epsilon^2}{||w| - u|}
\Big( {\partial_q^2} + {\partial_p^2} \Big)
+
\frac{w_q^2}{2} \; q^2  + \frac{w_p^2}{2} \; p^2
\eeq
with 
$V_1^\prime \equiv V(x_1^\prime,y_1^\prime)= u - |\omega| -{2\tau^2}/{(|\omega| - u)} -{UN}/{2}$.
Then the eigenvalues can be easily computed by considering the two independent harmonic-oscillator problems related to the coordinates $p$ and $q$, respectively.
$$
E_S(n,m)= V_1^\prime + 
\epsilon(|\omega| - u)\sqrt{1 - \frac{4\tau^2}{(|\omega| - u)^2}}\Big(n + \frac{1}{2}\Big)
$$
\beq
+ \epsilon(|\omega| - u) \sqrt{1 + \frac{4\tau^2(|\omega| + u)}{(|\omega| - u)^3}}
\Big (m + \frac{1}{2} \Big)\, .
\eeq
\end{appendix}

%%%%%%%%%%%%%%%%%%%%%%%%%%%%%%%%%%%%%%%%%%%%%%%%%%%%%%%%%%%%%%%%%%%%
%%%%%%%%%%%%%%%%%%%%%%%%%%%%%%%%%%%%%%%%%%%%%%%%%%%%%%%%%%%%%%%%%%%
%
%


\begin{thebibliography}{99}
%
% \bibitem{XXX} A.~C. Scott and J.~C.~Eilbeck, Phys. Lett. A {\bf 119}, 60 (1986).
% THE QUANTIZED DISCRETE SELF-TRAPPING EQUATION

\bibitem{rev1}
L.~Cruzeiro-Hansson, H.~Feddersen, R.~Flesch, P. L. Christiansen, M.~Salerno, and A.~C.~Scott, 
Phys. Rev. B {\bf 42},  522 (1990).
% Classical and quantum analysis of chaos in the DSTE

\bibitem{rev2}
E.~Wright, J.~C.~Eilbeck, M.~H. Hays, P.~D. Miller, and A.~C. Scott, Physica D {\bf 69}, 18 (1993).
% The quantum DSTE in the Hartree approximation
%

\bibitem{rev3} S. Flach and V. Fleurov, J. Phys.: Condens. Matter {\bf 9}, 7039 (1997).
% Tunnelling in the nonintegrable trimer—a step towards quantum breathers

\bibitem{rev4} A. Polkovnikov, S. Sachdev,  and S. M. Girvin, Phys. Rev. A {\bf 66}, 053607 (2002)
%

\bibitem{rev5} P. Buonsante, V. Penna, and A,. Vezzani, Phys. Rev. A {\bf 72}, 043620 (2005).
%
% \bibitem{rev2}
% F. Trimborn, D. Witthaut, and H. J. Korsch, Phys. Rev. A {\bf 79}, 013608 (2009).
% Beyond mean-field dynamics of small BH systems based on the number-conserving phase-space approach
%

% \bibitem{xxxxxxxxxxxx}
% R. A. Pinto and S. Flach, Phys. Rev. A {\bf 73},  022717 (2006)
% Quantum dynamics of localized excitations in a symmetric trimer molecule
%

\bibitem{rev6}
A. R. Kolovsky, H. J. Korsch, and E. M. Graefe, Phys. Rev. A {\bf 80},  023617 (2009).
% Bloch oscillations of Bose-Einstein condensates: Quantum counterpart of dynamical instability
%

\bibitem{jpa42}  P. Buonsante, R. Franzosi  and V. Penna, J. Phys. A {\bf 42}, 285307 (2009).

\bibitem{rev7} P. Buonsante, V. Penna, A. Vezzani, Phys. Rev. A {\bf 82}, 043615 (2010).
% quantum signatures of ST transition in attractive lattice bosons
% 

\bibitem{rev8}  
G. Mazzarella, L. Salasnich, A. Parola, and F. Toigo, Phys. Rev. A {\bf 83} 053607 (2011).
% Coherence and entanglement in the ground state of a bosonic Josephson junction: 
% From macroscopic Schrödinger cat states to separable Fock states
%

\bibitem{rev9} H. Hennig, D. Witthaut, and D. K. Campbell, Phys. Rev. A {\bf 86} 051604(R) (2012).
% quantum signatures of ST transition in attractive lattice bosons
% 

% \bibitem{rev8} P. J. Jason and M. Johansson, Phys. Rev. A 88, 033605 (2013).
%
\bibitem{rev10} P. J. Jason and M. Johansson, Phys. Rev. A {\bf 86}, 016214 (2012).
%

\bibitem{rev11} Xizhi Han and Biao Wu, Phys. Rev. A {\bf 93}, 023621 (2016).
% Ehrenfest breakdown of the mean-field dynamics of Bose gases
%

\bibitem{rev12} P. J. Jason and M. Johansson, Phys. Rev. E {\bf 94}, 052215 (2016).
% charge signatures of CFV in the BH trimer
%

\bibitem{sc1} L. Amico and V. Penna, Phys. Rev. Lett. {\bf 80}, 2189 (1998).
%

\bibitem{sc2} A. M. Rey, K. Burnett, R. Roth, M. Edwards, C. J. Williams and C. W. Clark,
J. Phys. B: At. Mol. Opt. Phys. {\bf 36}, 825 (2003).

\bibitem{sc3} J. Zakrzewski, Phys. Rev. A {\bf 71}, 043601 (2005).
% charge signatures of CFV in the BH trimer
%

% G. Mazzarella, L. Salasnich, A. Parola, and F. Toigo, 
% J. Phys. B: At. Mol. Opt. Phys. 43 (2010) 065303
% Nonlinear quantum model for atomic Josephson junctions with one and two bosonic species

%%%%%%%%%%%%%%%%%
\bibitem{cvp1} R. W. Spekkens and J. E. Sipe, Phys. Rev. A {\bf 59}, 3868 (1999).
% Spatial fragmentation of a Bose-Einstein condensate in a double-well potential
% Energy EV equation reduced to a harmonic-oscillator equation

% \bibitem{cvp2} R. Franzosi, V. Penna, and R. Zecchina, Phys. Rev. A {\bf }, 043609 (2001)
% Energy EV equation reduced to a harmonic-oscillator equation

\bibitem{cvp3} R. Franzosi and V. Penna, Phys. Rev. A {\bf 63}, 043609 (2001)
% Energy EV equation reduced to a harmonic-oscillator equation

\bibitem{cvp4} T.-L. Ho, C. V. Ciobanu, J. Low Temp. Phys. {\bf 135}, 257 (2004).
% The Schrödinger Cat Family in Attractive Bose Gases

\bibitem{cvp5} P. Zin, J. Chwedenczuk, B. Oles, K. Sacha, and M. Trippenbach,
Europhys. Letters {\bf 83}, 64007 (2008).
% Critical fluctuations of an attractive Bose gas in a double-well potential
% Energy EV equation reduced to a harmonic-oscillator equation

\bibitem{cvp6} P. Buonsante, R. Burioni, E. Vescovi, and A. Vezzani, Phys. Rev. A {\bf 85}, 043625 (2012).
% Quantum criticality in a bosonic Josephson junction
% Energy EV equation reduced to a harmonic-oscillator equation

\bibitem{cvp7} V. S. Shchesnovich and V. V. Konotop, Phys. Rev. A {\bf 75}, 063628 (2007).
% Nonlinear tunneling of Bose-Einstein condensates in an optical lattice:
% Signatures of quantum collapse and revival
% >>>> Schr. equation for a fictitious quantum particle in the one-dimensional discrete space

\bibitem{cvp8} J. Javanainen, Phys. Rev. A {\bf 60}, 4902 (1999).
% Phonon approach to an array of traps containing Bose-Einstein condensates

%%% \bibitem{jpa42}  P. Buonsante, R. Franzosi  and V. Penna, J. Phys. A {\bf 42}, 285307 (2009).

\bibitem{cvp9} P. Buonsante, V. Penna, A. Vezzani, Phys. Rev. A {\bf 84}, 061601(R) (2011).
% effective Hamiltonian
%
% \bibitem{kerdyk} R. J. Kerkdyk and S. Sinha, J. Phys. B: At. Mol. Opt. Phys. {\bf 46}, 185301 (2013).
% Effective potential method
% duale al CVP !!! il potenziale dipende da un angolo mentre N ne è la derivata

\bibitem{cvp10} G. Mazzarella, and V. Penna, J. Phys. B: At. Mol. Opt. Phys. {\bf 48}, 065001 (2015).
% Localization–delocalization transition of dipolar bosons in a four-well potential

%%%%%%%%%%%%%%%%%%%%%%%%%%%%%%%%%%%%%%%%%%%%%%%%%%%%%%%%%%%%%%%%%%%%%%%%%%

\bibitem{Em} C. Emary and T. Brandes, Phys. Rev. E {\bf 67}, 066203 (2003).

\bibitem{Fe} S. Felicetti, J. S. Pedernales, I. L. Egusquiza, G. Romero, L. Lamata,
D. Braak, and E. Solano, Phys. Rev. A {\bf 92}, 033817 (2015)

\bibitem{PeRa} V. Penna and F. A. Raffa, Int. J. Quantum Inform. {\bf 12}, 1560010 (2014).
%
% \bibitem{JJK} P. Jason, M. Johansson, and K. Kirr, Phys. Rev. E {\bf 86}, 016214 (2012).

\bibitem{VP} V. Penna, Phys. Rev. E {\bf 87}, 052909 (2013).

\bibitem{LMP} F. Lingua, G. Mazzarella, and V. Penna, J. Phys. B {\bf 49}, 205005 (2016).

%%%%%%%%%%%%%%%%%%%%%%%%%%%%%%%%%%%%%%%%%%%%%%%%%%%%%%%%
% \cite{ashhab} S. Ashhab and C. Lobo, Phys. Rev. A 66 (2002) 013609

\bibitem{xu} X. Q. Xu, L. H. Lu, and Y. Q. Li, Phys. Rev. A 78 (2008) 043609

\bibitem{satjia} I. I. Satija, R. Balakrishnan, P. Naudus, J. Heward, M. Edwards and C. W. Clark, Phys. Rev. A 79 (2009) 033616

\bibitem{juliad} B. Juli\'a-D\'iaz, 
M. Mel\'e-Messeguer, M. Guilleumas and A. Polls, Phys. Rev. A 80 (2009) 043622

\bibitem{mazz2011} G. Mazzarella, B. Malomed, L. Salasnich, M. Salerno and F. Toigo, J. Phys. B: At. Mol. Opt. Phys. 44 (2011) 035301
% Rabi–Josephson oscillations and self-trapped dynamics in atomic junctions with two bosonic species

\bibitem{citro} A. Naddeo and R. Citro, J. Phys. B: At. Mol. Opt. Phys. 43 (2010) 135302

% \bibitem{zin} P. Zi\'n P, B. Ole\'s and K. Sacha, Phys. Rev. A 84 (2011) 033614

\bibitem{mujal} P. Mujal, B. Jul\'ia-D\'iaz, and A. Polls, Phys. Rev. A {\bf 93}, 043619 (2016) 
% Quantum properties of a binary bosonic mixture in a double well

\bibitem{baizakov} B. B. Baizakov, A. Bouketir, A. Messikh, and B. A. Umarov, Phys. Rev. E 79 (2009) 046605

\bibitem{huang} J.-S. Huang, Z.-W. Xie, M. Zhang and L.-F. Wei, J. Phys. B: At. Mol. Opt. Phys. 43 (2010) 065305

\bibitem{kevre} P. G. Kevrekidis, {\it The Discrete Nonlinear Schr\"odinger Equation}, (Springer-Verlag Berlin 2009) 

\bibitem{gutz} 
% M.  C.  Gutzwiller, J.  Math.  Phys. 12,  343  (1971);  
M. C. Gutzwiller, {\it Chaos  in  Classical  and  Quantum  Mechanics}, (Springer-Verlag, New York, 1990). 

\bibitem{landau}
L. D. Landau L D and E. M. Lifsits, {\it Quantum Mechanics} (Pergamon, Oxford, 1957)

\bibitem{pra92} F. Lingua, M. Guglielmino, V. Penna, and B. Capogrosso Sansone,
Phys. Rev. A {\bf 92}, 053610 (2015).

\bibitem{atom1} D. Aghamalyan, L. Amico, and L. C. Kwek, Phys. Rev. A {\bf 88}, 063627 (2013)
% Effective dynamics of cold atoms flowing in two ring-shaped optical potentials with tunable tunneling

\bibitem{atom2} L. Amico, D. Aghamalyan, F. Auksztol, H. Crepaz, R. Dumke, and L. C. Kwek, 
Sci. Rep. {\bf 4}, 4298 (2014)
% Superfluid qubit systems with ring shaped optical lattices

\bibitem{atom3} M. K. Olsen, and J. F. Corney, Phys. Rev. A {\bf 94}, 033605 (2016).

\bibitem{atom4} R. A. Pepino, J. Cooper, D. Meiser, D. Z. Anderson, and M. J. Holland,
Phys. Rev. A {\bf 82}, 013640 (2010)
% Open quantum systems approach to atomtronics

\bibitem{atom5} R. Mathew, A. Kumar, S. Eckel, F. Jendrzejewski, G. K. Campbell, 
M. Edwards, and E. Tiesinga, Phys. Rev. A {\bf 92}, 033602 (2015).
%Self-heterodyne detection of the in situ phase of an atomic superconducting quantum interference device

%
\bibitem{albiez} Albiez M, Gati R, Fölling J, Hunsmann S, Cristiani M and Oberthaler M K 2005 Phys. Rev. Lett. 95 010402
%
\bibitem{anker} Th. Anker, M. Albiez, R. Gati, S. Hunsmann, B. Eiermann, A. Trombettoni, and M. K. Oberthaler, Phys. Rev. Lett. 94, 020403 (2005).
%
\bibitem{lens} J. Catani, L.De Sarlo,G. Barontini, F. Minardi, and M. Inguscio, Phys. Rev. A 77, 011603 (2008).
%
\bibitem{gadway} B. Gadway, D. Pertot, R. Reimann, and D. Schneble, Phys. Rev. Lett. 105, 045303 (2010).
%


% B. T. Seaman, M. Krämer, D. Z. Anderson, and M. J. Holland,  Phys. Rev. A {\bf 75}, 023615 (2007)
% Atomtronics: Ultracold-atom analogs of electronic devices

% R. A. Pepino, J. Cooper, D. Meiser, D. Z. Anderson, and M. J. Holland, Phys. Rev. A {\bf 82}, 013640 (2010)
% Open quantum systems approach to atomtronics

% A Gallemí, M Guilleumas, J Martorell, R Mayol, A Polls and B Juliá-Díaz, 
% New J. Phys. 17 (2015) 073014
% Fragmented condensation in Bose–Hubbard trimers with tunable tunnelling

% D. Aghamalyan, M. Cominotti, M. Rizzi, D. Rossini, F. Hekking, A. Minguzzi, L-C Kwek and L. Amico, 
% New J. Phys. 17, 045023 (2015)

% R. Mathew, A. Kumar, S. Eckel, F. Jendrzejewski, G. K. Campbell, Mark Edwards, and E. Tiesinga
% Phys. Rev. A {\bf 92}, 033602 (2015)
% Self-heterodyne detection of the in situ phase of an atomic superconducting quantum interference device

% A. A. Zozulya, D. Z. Anderson, Phys. Rev. A {\bf 88}, 043641 (2013)
% Principles of an atomtronic battery

% M. K. Olsen, A. S. Bradley, Phys. Rev. A {\bf 91}, 043635 (2015)
% Quantum ultracold atomtronics

\end{thebibliography}
\end{document}